%% file: main.tex
\pdfoutput=1 
\documentclass[10pt, journal]{IEEEtran}
\newcommand{\ignore}[1]{}
\usepackage[boxed]{algorithm}
\usepackage{fancyhdr}
\usepackage[normalem]{ulem}
\usepackage[hyphens]{url}
\usepackage{microtype}
\usepackage{graphicx}
\usepackage{xcolor}
\usepackage{xspace}
\usepackage{multirow}
\usepackage{multicol}
\usepackage{authblk}
\usepackage{graphicx}
\usepackage{array}
\usepackage{booktabs}
\usepackage{amsmath}
\usepackage{amssymb}
\usepackage{amsfonts}
\usepackage{amsthm}
\usepackage{supertabular}
\usepackage{bm}
\newcommand{\eat}[1]{}
\usepackage[T1]{fontenc}
\usepackage{subfigure}
\usepackage[utf8]{inputenc}
\usepackage{colortbl} 
\usepackage{arydshln} 
\usepackage{threeparttable}
\usepackage{algorithm}
\usepackage{algorithmic}
\usepackage{amsbsy}
\usepackage{color}

\begin{document}

\title{RCoNet: Deformable Mutual Information Maximization and High-order Uncertainty-aware Learning for Robust COVID-19 Detection}

\author{Shunjie~Dong,~\IEEEmembership{}
       Qianqian~Yang,~\IEEEmembership{} 
       Yu~Fu,~\IEEEmembership{}
        Mei~Tian,~\IEEEmembership{}
        Cheng~Zhuo,~\IEEEmembership{Senior~Member,~IEEE}
        % \thanks{Manuscript received May 27, 2019; revised September 30, 2019. This paper was recommended by Associate Editor A. Coskun.
        % This work was supported in part by the NSFC under grant 61974133 and 61601406, Zhejiang Provincial Key R\&D program under grant 2020C01052, National Key R\&D Program of China under grant 2018YFE0126300, and in part by the NRI Program of SRC, by the NSF under grant 1640081, and EXCEL, an SRC-NRI Nanoelectronics Research Initiative under Research Task ID 2698.004, and Asian Research Grant from the University of Notre Dame.
        % Date of publication xx xx, xxxx; date of current version xx xx, xxxx. \textit{(Corresponding author: Cheng Zhuo.)}}
        % \thanks{D. Gao and C. Zhuo are with the department of Information Science \& Electronic Engineering, Zhejiang University, Hangzhou 310027, China (e-mail: czhuo@zju.edu.cn).}
        % \thanks{D. Reis and X. S. Hu are with the University of Notre Dame, Notre Dame, IN 46556, USA.}
        % \thanks{Digital Object Identifier }

        }% <-this % stops a space

\maketitle
\begin{abstract}
The novel 2019 Coronavirus (COVID-19) infection has spread world widely and is currently a major healthcare challenge around the world. Chest Computed Tomography (CT) and X-ray images have been well recognized to be two effective techniques for clinical COVID-19 disease diagnoses. Due to faster imaging time and considerably lower cost than CT, detecting COVID-19 in chest X-ray (CXR) images is preferred for efficient diagnosis, assessment and treatment. However, considering the similarity between COVID-19 and pneumonia, CXR samples with deep features distributed near category boundaries are easily misclassified by the hyper-planes learned from limited training data. Moreover, most existing approaches for COVID-19 detection focus on the accuracy of prediction and overlook the uncertainty estimation, which is particularly important when dealing with noisy datasets. To alleviate these concerns, we propose a novel deep network named {\em RCoNet$^k_s$} for robust COVID-19 detection which employs {\em Deformable Mutual Information Maximization} (DeIM), {\em Mixed High-order Moment Feature} (MHMF) and {\em Multi-expert Uncertainty-aware Learning} (MUL). With DeIM, the mutual information (MI) between input data and the corresponding latent representations can be well estimated and maximized to capture compact and disentangled representational characteristics. Meanwhile, MHMF can fully explore the benefits of using high-order statistics and extract discriminative features of complex distributions in medical imaging. Finally, MUL creates multiple parallel dropout networks for each CXR image to evaluate uncertainty and thus prevent performance degradation caused by the noise in the data. The experimental results show that RCoNet$^k_s$ achieves the state-of-the-art performance on an open source COVIDx dataset of 15134 original CXR images across several metrics. Crucially, our method is shown to be more effective than existing methods with the presence of noise in the data.
\end{abstract}


\IEEEpeerreviewmaketitle

\input{introduction.tex}
\input{related_work.tex}
\input{method.tex}
\input{results.tex}

\section{Conclusions}
\label{sec_conclusion}
In this paper, we proposed a novel deep network model, named {\em RCoNet$^k_s$}, for robust COVID-19 detection, which contains three key components, i.e., {\em Deformable mutual Information Maximization} (DeIM), {\em Mixed High-order Moment Feature} (MHMF) and {\em Multi-expert Uncertainty-aware Learning} (MUL). DeIM estimates and maximizes the mutual information between input data and the latent representations simultaneously to obtain the category separability in the latent space. We proposed MHMF to overcome the limited expressive capability of low-order statistics, and instead use a combination of both low and high order moment features to extract more informative and discriminative features. MUL generates the final diagnosis and the uncertainty estimation, by combining the output of multiple parallel dropout networks, each as an expert. We numerically validated that the proposed RCoNet trained on either the public COVIDx dataset or the noisy version of it, outperforms the existing methods in terms of all the metrics considered. We note that these three modules can be easily implemented into other frameworks for different tasks.

%%%%%%%%%%%%%%%%%%%%%%%%%%%%%%%%%%%%%%%%%%%%%%%%%%%%%%%%%%%%%%%%%%%%%%%%%%%%%%%%%%%%%%%%%%%%%%%%%%%%%%%%%%%%%%%%%%%%%%%%%
\bibliographystyle{IEEEtran}
\bibliography{ref}

\end{document}

%% file: introduction.tex
\section{Introduction}
\label{sec_introduction}

\IEEEPARstart{C}{}ORONAVIRUS disease 2019 (COVID-19) causes an ongoing pandemic that significantly impacts everyone's life since it was first reported, with hundreds of thousands of deaths and millions of infections emerging in over 200 countries~\cite{zhang2020clinically,han2020accurate}. As indicated by the World Health Organization (WHO), due to its highly contagious nature and lack of corresponding vaccines, the most effective method to control the spread of COVID-19 infection is to keep social distance and contact tracing. Hence, early and fast diagnosis of COVID-19 has become significantly essential to control further spreading, and such that the patients could be hospitalized and receive proper treatment in time.

Since the emerge of COVID-19, {\em reverse transcription polymerase chain reaction} (RT-PCR), as a viral nucleic acid detection method by gene sequencing, is the accepted standard for COVID-19 detection~\cite{mei2020artificial}. However, because of the low accuracy of RT-PCR and limited medical test kits in many hyper-endemic regions or countries, it is challenging to detect every individual affected by COVID-19 rapidly~\cite{xie2020relational,ouyang2020dual}. Therefore, alternative testing methods, which are faster and more reliable than RT-PCR, are urgently needed to combat the disease.
\begin{figure}[tb]
    \centering 
    \includegraphics[width=8.5cm]{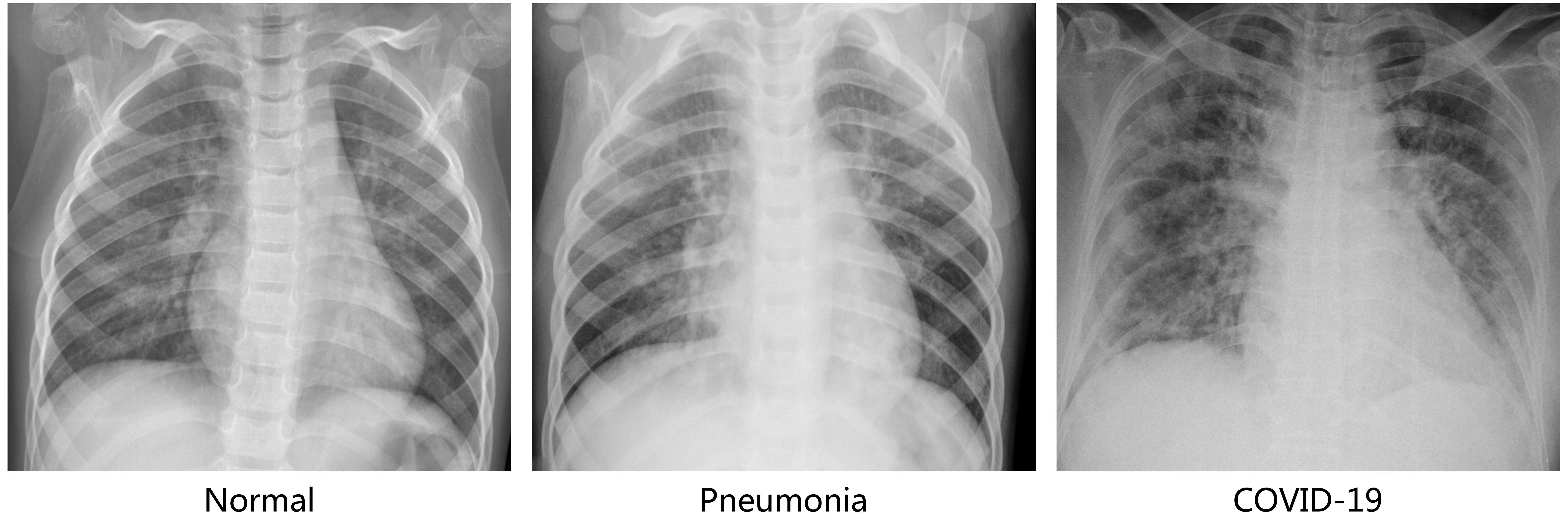}
    \caption{Visual illustration of chest X-ray images, including normal, pneumonia and COVID-19.}
    \label{sample}
\end{figure}

Since most COVID-19 positive patients were diagnosed with pneumonia, radiological examinations could help detect and assess the disease. Recently, chest {\em computed tomography} (CT) has been shown to be efficient and reliable to achieve a real-time clinical diagnosis of COVID-19, outperforming over RT-PCR in terms of accuracy. Moreover, some deep learning based methods have been proposed for COVID-19 detection using chest CT images~\cite{bai2020ai,ardakani2020application,kang2020diagnosis,fan2020inf}. For example, an adaptive feature selection approach was proposed in~\cite{sun2020adaptive} for COVID-19 detection based on a trained deep forest model. In~\cite{donglin2020hypergraph}, an uncertainty vertex-weighted hypergraph learning method was designed to identify COVID-19 from {\em community acquired pneumonia} (CAP) using CT images. However, the routine use of CT, which is conducted via expensive equipments, takes considerably more time than X-ray imaging and brings a massive burden on radiology departments. Compared to CT, X-rays could significantly speed up disease screening, and hence become a preferred method for disease diagnosis.

Accordingly, deep learning based methods for detecting COVID-19 with {\em chest X-ray} (CXR) have been developed and shown to be able to achieve accurate and speedy detection~\cite{zu2020coronavirus,siddhartha2020covidlite}. For instance, a tailored convolution neural network platform trained on open source dataset called COVIDNet in~\cite{wang2020covid} was proposed for the detection of COVID-19 cases from CXR. Oh \textit{et al.}~\cite{oh2020deep} proposed a novel probabilistic gradient-weighted class activation map to enable infection segmentation and detection of COVID-19 on CXR images. Fig.~\ref{sample} shows three samples from the {\em COVIDx} dataset~\cite{wang2020covid} which contains three different classes: normal, pneumonia and COVID-19. However, due to the similar pathological information between pneumonia and COVID-19 in the early stage, the CXR samples may have latent features distributed near the category boundaries, which can be easily misclassified by the hyper-plane learned from the limited training data. Moreover, to the best of our knowledge, most of the existing methods for COVID-19 detection are designed to extract the lower-dimension latent representations which may not be able to fully capture statistical characteristic of complex distributions (i.e., non-Gaussian distribution). Furthermore, quantifying uncertainty in COVID-19 detection is still a major yet challenging task for doctors, especially with the presence of noise in the training samples (i.e., {\em label} noise and {\em image} noise).

To address the above problems, we propose a novel deep network architecture, referred to as {\em RCoNet$^k_s$}, for robust COVID-19 detection which, in particular, contains the following three modules, i.e.,  {\em Deformable mutual Information Maximization} (DeIM), {\em Mixed High-order Moment Feature} (MHMF) and {\em Multi-expert Uncertainty-aware Learning} (MUL): 
\begin{itemize}
    \item The Deformable mutual Information Maximization (DeIM) module estimates and maximizes the mutual information (MI) between input data and learned high-level representations, which pushes the model to learn the discriminative and compact features. We employ deformable convolution layers in this module which are able to explore disentangled spatial features and mitigate the negative effect of similar samples across different categories. 
    \item The Mixed High-order Moment Feature (MHMF) module, inspired by~\cite{pauwels2016sorting}, fully explores the benefits of using a mix of high-order moment statistics to better characterize the feature distributions in medical imaging.
    \item The Multi-expert Uncertainty-aware Learning (MUL) creates multiple parallel dropout networks, each can be treated as an {\em expert}, to derive multiple experts based diagnosis similar to clinical practices, which improves the prediction accuracy. MUL also quantifies the prediction accuracy by obtaining the variance in prediction across different experts.
    \item The experimental results show that our proposal achieves the state-of-the-art performance in terms of most metrics both on open source COVIDx dataset of 15134 original CXR images and that of noisy setting.
\end{itemize}

The remaining of this paper is organized as follows: In Section~\ref{sec_background}, we review related works on mutual information estimation and uncertainty learning as well. In Section~\ref{sec_method}, after an overview of our proposed approach, we discuss the main components of RCoNet$^k_s$. In Section~\ref{sec_experiment}, we compare our proposed architecture with the existing deep learning based methods evaluated on a public available dataset of CXR images and also the same dataset but under noisy conditions. And we also conduct extensive experiments to demonstrate the benefits of DeIM, MHMF and MUL on the performance of the system. Finally, we conclude this paper in Section~\ref{sec_conclusion}.

%% file: related_work.tex
\section{Background and Related Works}
\label{sec_background}

In this section, we introduce related works on mutual information estimation and uncertainty learning that lay the foundation of this paper.

\subsection{Mutual Information Estimation}
Mutual information (MI), as a fundamental concept in information theory, is widely applied to unsupervised feature learning for quantifying the correlation between random variables. MI has been exploited in a wide range of domains and tasks, including biomedical sciences~\cite{maes1997multimodality}, blind source separation (BSS, e.g., independent component analysis~\cite{hyvarinen2000independent}), feature selection~\cite{kwak2002input,peng2005feature} and causal inference~\cite{butte1999mutual}. For example, the object tracking task considered in~\cite{vondrick2018tracking} was treated as a problem of optimizing the mutual information between features extracted from a video with most color information removed and those from the original full-color video. Closely related work presented in~\cite{arandjelovic2017look} considered learning representations to predict cross-modal correspondence by maximizing MI between features from the multi-view encoders and the content of the held-out view. Moreover, Mutual Information Neural Estimation (MINE) proposed by~\cite{belghazi2018mutual} was designed to learn a general-purpose estimator of the MI between continuous variables based on dual representations of the KL-divergence, which are scalable, flexible and, most crucially, trainable via back-propagation. Based on MINE, our proposal estimates and maximizes the CXR image inputs and the corresponding latent representations to improve diagnosis performance.

\begin{figure*}[htb]
    \centering 
    \includegraphics[width=0.9\textwidth]{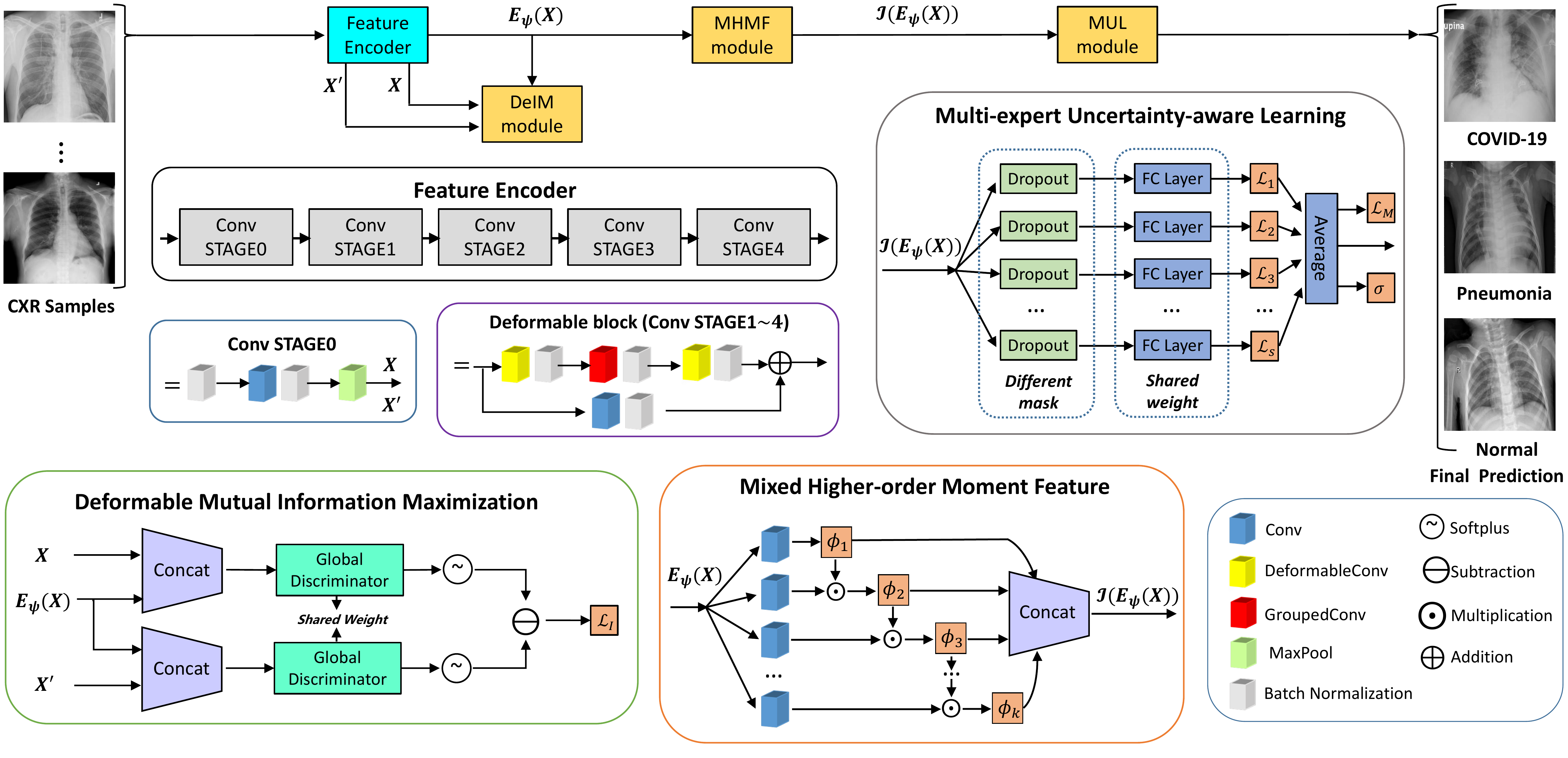}
    \caption{The architecture of RCoNet$^k_s$ for COVID-19 detection.}
    \label{framework}
\end{figure*}

\subsection{Uncertainty in Deep Learning}
Aiming at combating the significant negative effects of uncertainty in deep neural networks, uncertainty learning has been getting lots of research attention, which facilitates the reliability assessment and solves risk-based decision-making problems~\cite{chang2020data,shi2019probabilistic,kendall2015bayesian}. In recent years, various frameworks have been proposed to characterize the uncertainty in the model parameters of deep neural networks, referred to as \textit{model uncertainty}, due to the limited size of training data~\cite{blundell2015weight,gal2016uncertainty}, which can be reduced by collecting more training data~\cite{shi2019probabilistic,mackay1992practical,neal2012bayesian}. Meanwhile, another kind of uncertainty in deep learning, referred to as \textit{data uncertainty}, measures the noise inherent in given training data, and hence cannot be eliminated by having more training data~\cite{kendall2017uncertainties}.
To combat these two kinds of uncertainty, lots of works on various computer vision tasks, i.e., face recognition~\cite{chang2020data}, semantic segmentation~\cite{isobe2017deep}, object detection~\cite{choi2019gaussian}, person re-identification~\cite{yu2019robust}, etc., have introduced deep uncertainty learning to improve the robustness of deep learning model and interpretability of discriminant. For face recognition task in~\cite{shi2019probabilistic}, an uncertainty-aware probabilistic face embedding (PFE) was proposed to represent face images as distributions by utilizing data uncertainty. 
Exploiting the advantage of Bayesian deep neural networks, one recent study~\cite{zafar2019face} leveraged the model uncertainty for analysis and learning of face representations. 
To our knowledge, our proposal is the first work that utilizes the high-order moment statistics and multiple expert networks to estimate uncertainty for COVID-19 detection using CXR images. 

%% file: method.tex
\section{Method}
\label{sec_method}
In this section, we introduce the novel {\em RCoNet$^k_s$} for robust COVID-19 detection, which incorporates {\em Deformable mutual Information Maximization} (DeIM), {\em Mixed High-order Moment Feature} (MHMF) and {\em Multi-expert Uncertainty-aware Learning} (MUL), as illustrated in Fig.~\ref{framework}. $k$ is the number of levels of moment features that are combined in MHMF, and $s$ is the number of the expert network in MUL, which will be further clarified in the sequel. The CXR images are first processed by DeIM which consists of a stack of deformable convolution layers, extracting discriminative features. The compact features are then fed into MHMF module to generate high-order moment latent features, reducing negative effects caused by similar images. The proposed MUL utilizes the learned high-order features to generate final diagnoses.

\subsection{Deformable Mutual Information Estimation and  Maximization}
\label{subsec_DeIM}
Due to the similarity between COVID-19 and pneumonia in the latent space, we propose Deformable mutual Information Maximization (DeIM) to extract discriminative and informative features, reducing the negative influence caused by the lack of distinctiveness in the deep features. In particular, we train the model by maximizing the mutual information between the input and corresponding latent representation. 

We use a stack of five convolutional stages, as shown in Fig.~\ref{framework}, to encode inputs into latent representations, which is denoted by a differentiable parametric function $E_\psi$:
\begin{equation}
E_\psi:\mathcal{X}\rightarrow\mathcal{Z},
\end{equation}
where $\psi$ denotes the set of all the trainable parameters in these layers, and $\mathcal{X}$ and $\mathcal{Z}$ denote the input and output spaces, respectively. 

The detailed architecture of each convolutional stage is presented in Fig.~\ref{framework}, which consists of several convolutional layers each followed by a batch normalization layer. Note that we employ deformable convolutional layers which can better extract spatial information of the irregular infected area compared to conventional convolutional layers. More specifically, regular convolution operates on pre-defined rectangular grid from an input image or a set of input feature maps, while the deformable convolution operates on deformable grids that each grid point is moved by a learnable offset. For example, the receptive grid $\mathcal{P}$ of a regular convolution with kernel size $3\times3$ is fixed and can be given by:
\begin{equation}
\mathcal{P}=\{(-1,-1),(-1,0),...,(0,1),(1,1)\},
\end{equation}
while, for deformable convolution, the receptive grid is moved by the learned offsets $\Delta p_n \in \mathbb{R}^{2}$ and the output is given as follows:
\begin{equation}
b(p_0)=\sum_{P_n\in \mathcal{P}}w(p_n)\cdot a(p_0+p_n+\Delta p_n).
\end{equation}
where $b(p_0)$ denotes the value at location $p_0$ on the output feature map $b$, $p_n$ enumerates the locations in $\mathcal{P}$,  $w(p_n)$ represents the weight at location $p_n$ of the kernel, and $a(\cdot)$ is value at given location on the input feature map. We can see that with the introduction of offsets $\Delta p_n$, the receptive grid is no longer fixed to be a rectangle, and instead is deformable.

We optimize $E_\psi$ by maximizing the mutual information between the input and the output, i.e., $I(X;Z)$, where $Z\triangleq E_\psi(X)$. The precise mutual information requires knowledge probability density functions (PDFs) of $X$ and $Z$, which is intractable to obtain in practice. To overcome this issue, Mutual Information Neural Estimation (MINE) proposed in \cite{belghazi2018mutual} estimates mutual information by using a lower-bound on the Donsker-Varadhan representation \cite{donsker1975asymptotic} of the KL-divergence: 
\begin{equation}
\begin{aligned}
I(X;Z):&=D_{KL}(\mathbb{J}||\mathbb{M})\geq \widehat{I}_\theta^{(DV)}(X;Z):\\
&=\mathbb{E}_\mathbb{J}[T_\theta(x,z)]-\log\mathbb{E}_\mathbb{M}[e^{T_\theta(x,z)}],
\label{dv}
\end{aligned}
\end{equation}
where $\mathbb{J}$ represents the joint probability of $X$ and $Z$, i.e., $\mathbb{J}\triangleq P(X,Z)$, and $\mathbb{M}$ denotes the product of marginal probabilities of $X$ and $Z$, $\mathbb{M}\triangleq P(X)P(Z)$.  $T_\theta:\mathcal{X}\times\mathcal{Z}\rightarrow\mathbb{R}$ denotes a \textit{global discriminator} modeled by a neural network with parameters $\theta$, which is trained to maximize $\widehat{I}_\theta^{(DV)}(X;Z)$ to approximate the actual mutual information. Hence, we can simultaneously estimate and maximize $I(X;E_{\psi}(X))$ by maximizing $\widehat{I}_\theta^{(DV)}(X;Z)$:
\begin{equation}
\begin{aligned}
(\widehat{\theta},\widehat{\psi})&=\underset{\theta,\psi}{arg  max}\widehat{I}^{(DV)}_\theta(X;E_\psi(X)).
\label{mi}
\end{aligned}
\end{equation}
Since the encoder $E_{\psi}$ and the mutual information estimator $T_\theta$ are optimized simultaneously with the same objective function, we can share some layers between them, and replace the $T_\theta$ with $T_{\theta, \psi}$ to account for this fact.

Since we are primarily interested in maximizing the mutual information rather than estimating the precise value, we can alternatively use a Jensen-Shannon MI estimator (JSD)~\cite{nowozin2016f}, which offers more interpretable trade-off:
\begin{equation}
\begin{aligned}
\widehat{I}_{\theta,\psi}^{(DeJSD)}(X;E_\psi(X)&):=\mathbb{E}_\mathbb{P}\left[-\log\left(1+e^{-T_{\theta,\psi}(x,E_\psi(x))}\right)\right]\\
&-\mathbb{E}_{\mathbb{P}\times\widetilde{\mathbb{P}}}\left[\log\left(1+e^{T_{\theta,\psi}(x',E_\psi(x))}\right)\right],
\label{loss_jsd}
\end{aligned}
\end{equation}
where $x$ is an input sample of an empirical probability distribution $\mathbb{P}$, $x'$ denotes a fake sample from distribution $\widetilde{\mathbb{P}}$, where $\widetilde{\mathbb{P}} = \mathbb{P}$. 
This estimator is illustrated by th DeIM block shown in Fig.~\ref{framework}, which has the latent representation $E_\psi(x)$, the input sample $x$ and the fake sample $x'$ as input, and the difference between the outputs of the two softplus operations as the estimation of MI.

Another alternative MI estimator is called Noise-Contrastive Estimator (NCE) \cite{gutmann2012noise}, which is defined as:
\begin{equation}
\begin{aligned}
\widehat{I}&_{\theta,\psi}^{(DeNCE)}(X;E'_\psi(X)):=\\&\mathbb{E}_\mathbb{P}\left[T_{\theta,\psi}(x,E'_\psi(x))
-\mathbb{E}_{\widetilde{\mathbb{P}}}\left[\log\sum_{x'} e^{T_{\theta,\psi}(x',E'_\psi(x))}\right]\right].
\label{nce}
\end{aligned}
\end{equation}
The experiments have found that using the NCE estimator outperforms the JSD estimator in some cases, but appears to be quite similar most of the time.

The existing works~\cite{bachman2019learning} that implement these estimators use some latent representation of $x$, which is then merged with some randomly generated features to obtain ``fake'' samples that satisfy $\mathbb{P}=\widetilde{\mathbb{P}}$. In contrast, we use the samples from other categories as the ``fake'' samples, i.e., $x'$, instead. For example, if the input is a pneumonia sample, then the fake sample is either a normal or COVID sample. We note that this can push the learned encoder to derive more distinguishable features for samples from different categories. 

%%%%%%%% begin figure 3 %%%%%%%%%%
\begin{figure}[tb]
    \centering 
    \includegraphics[width=8.7cm]{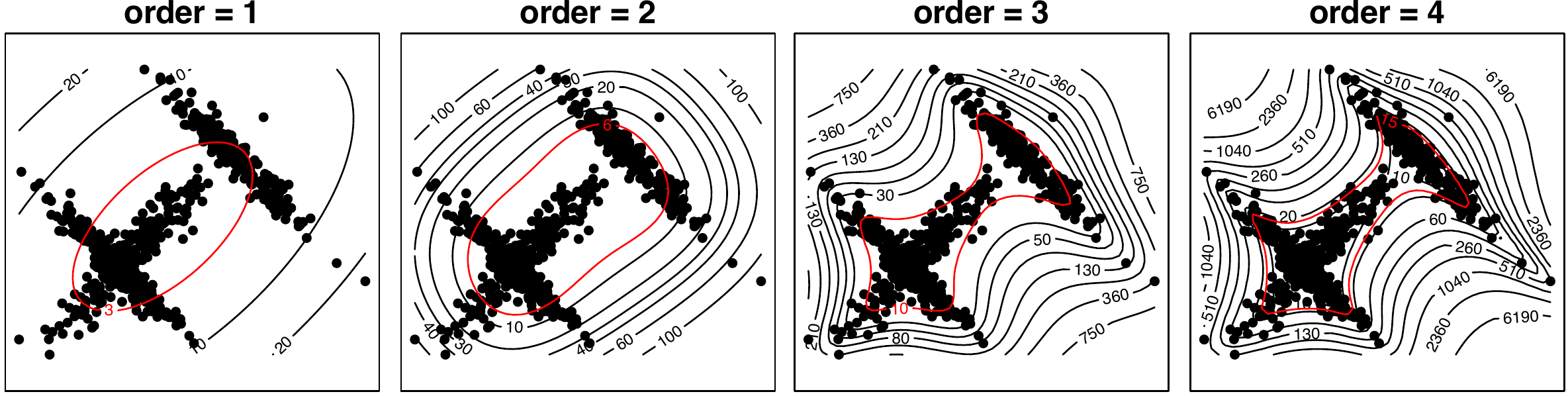}
    \caption{Data points from three Gaussian distributions and the corresponding moment feature of order 1 to 4}.
    \label{higher_order}
\end{figure}
%%%%%%%% end figure 3 %%%%%%%%%%
\subsection{Mixed High-order Moment Feature}
\label{subsec_MHMF}
The presence of the image noise and label noise in CXR datasets may cause image latent representations generated by deep neural networks to be scattered in the entire feature space. To deal with this issue, \cite{chang2020data,shi2019probabilistic,yu2019robust} represent each image as a Gaussian distribution, that is defined by a mean (a standard feature vector) and a variance. However, the deep features of CXR samples we considered in this paper typically follow a complex, non-Gaussian distribution~\cite{xu2016blind,chen2019homm}, which cannot be fully captured by its first-order (mean) or second-order statistics (variance).

We seek a better combination of different orders of statistics to more precisely characterize the latent representation of the CXR images. We illustrate the moment features of different orders \cite{pauwels2016sorting} in Fig.~\ref{higher_order}, where we plot 350 data points in $\mathbb{R}^2$ sampled from a distribution that combines three different Gaussian distributions. We can observe that the high-order moment features are more expressive of statistical characteristic compared to low-order one. More specifically, it captures the shape of the cloud of samples more accurately. 
Therefore, we include the Mixed High-order Moment Feature (MHMF) module in the proposed model, as shown in Fig.~\ref{framework}, which outputs a combination of high-order moment features with the latent representation $E_\psi(X)$ as input. This will potentially solve the scattering problem, and, more importantly, capture the subtle differences between CXR images of similar categories, i.e., pneumonia and COVID-19 in our case.

We show how to obtain the complicated high-order moment feature in the following. Define $r$-th order moment feature as $\phi_{r}(\bm{a})$, where $\bm{a}\in \mathbb{R}^{H\times W\times C}$ denotes a latent feature map of dimension $H\times W\times C$. Lots of recent works adopt the Kronecker product to compute high-order moment feature~\cite{chen2019homm}.
However, calculating Kronecker product of high dimensional feature maps is significantly computational intensive, and hence infeasible for real-world applications. Inspired by \cite{jacob2019metric,jegou2012negative,opitz2017bier}, we approximate $\phi_{r}(\bm{a})$ by exploiting $r$ random projectors which relies on certain factorization schemes, such as Random Maclaurin~\cite{kar2012random}. We use $1\times1$ convolution kernels as the random projectors to estimate the expectations of high-order moment features. That is, 
\begin{equation}
\label{loss_in}
  \phi_{r}(\bm{a})\approx \mathcal{K}_1(\bm{a})\odot\mathcal{K}_2(\bm{a})\odot\cdots\odot\mathcal{K}_r(\bm{a}) \in \mathbb{R}^{H \times W \times C},
\end{equation}
where $\odot$ represents the Hadamard (element-wise) product, and $\mathcal{K}_1,\mathcal{K}_2,\dots,\mathcal{K}_r$ are $1\times1$ convolution kernels with random weights.

Note that Random Maclaurin produces a estimator that is independent of the input distribution, which causes the estimated high-order moments to contain non-informative high-order moment components. We eliminate these components by learning the weights of the projectors, i.e., the $1\times1$ convolution kernels, from the data. Also note that the Hadamard product of a number of random projectors may end up with the estimated high-order moment features to be similar to low-order ones. To solve this problem, we use a recursive way to estimate the high-order moments instead,
\begin{equation}
\label{loss_r}
  \phi_r(\bm{a}) = \phi_{r-1}(\bm{a}) \odot \mathcal{K}_r(\bm{a}).
\end{equation}

Since different order moments capture different informative statistics, we design the MHMF module to keep the estimated moments of different levels of order, as shown in Fig.~\ref{framework}, the output of which is given as:
\begin{equation}
\label{loss_MHMF}
  \mathcal{J}(\bm{a}) = [\phi_1(\bm{a});\phi_2(\bm{a});\cdots;\phi_r(\bm{a})]\in\mathbb{R}^{H \times W \times rC}.
\end{equation}
Hence, $\mathcal{J}(\bm{a})$ is rich enough to capture the complicated statistics, and produce discriminative features for the input of different categories.

\subsection{Multi-expert Uncertainty-aware Learning}
\label{subsec_MUL}
The MHMF module, as described in section \ref{subsec_MHMF}, generates mixed high-order moment features of each sample in the latent space, which we aim to further exploit to derive compact and disentangled information for COVID-19 detection. 
Meanwhile, quantifying uncertainty in disease detection is undoubtedly significant to understand the confidence level of computer-based diagnoses. 
Motivated by the clinical practices, we present a novel neural network in this section, referred to as Multi-expert Uncertainty-aware Learning (MUL), which takes in the mixed high-order moment features and outputs the prediction and the quantification of the diagnostic uncertainty caused by the noise in the data.

The structure of Multi-expert Uncertainty-aware Learning module is shown in Fig.~\ref{framework}, which consists of multiple dropout layers that process the output from MHMF in parallel, each of which together with the following several fully connected layers can be regarded as an {\em expert} for COVID-19 detection. We note that each dropout layer uses different masks which results in different subsets of latent information to be kept, while the following fully connected layers share the same weights across different experts. The masks for the dropout layers are generated randomly at each iteration during training, but fixed during the inference time.  
We denote the input-output function of each expert by $C_{e}^j(\cdot)$, $j= 1, ..., N$, where $N$ is the total number of experts.  Hence, we have the classification loss $\mathcal{L}_e^j$ of $j$-th expert given as follows:
\begin{equation}
\label{loss_softmax}
  \mathcal{L}_e^j = \frac{1}{n}\sum^{n}_{i=1}\mathcal{L}_{w}(C_{e}^j(\mathcal{J}(E_\psi(x_i))),y_i),
\end{equation}
where $n$ represents the total number of labeled CXR samples, and $y_i$ denotes the one-hot representation of the class label, $i = 1, ..., n$, and we recall that $\mathcal{J}(\cdot)$ denotes the MHMF operation given in Eq.~\eqref{loss_MHMF} and $E_\psi(\cdot)$ is the preprocessing step on the CXR samples. 
Note that, the total number of COVID-19 cases is much smaller than non-COVID cases, i.e., normal and pneumonia cases. This imbalance in the dataset leads to a high ratio of false-negative classification. To mitigate this negative effect, we employ a weighted cross-entropy $\mathcal{L}_{w}(\cdot)$ given as follows:
\begin{equation}
\label{loss_weight}
  \mathcal{L}_w(\widehat{y}_i, y_i) = -\frac{1}{C}\sum^{C}_{c=1}\lambda_c\cdot y_{i, c}\log\widehat{y}_{i, c},
\end{equation}
where $C$ is the total number of classes, $y_{i, c}$ is the $c$-th element of $y_i$, and $\widehat{y}_{i, c}$ denotes the corresponding prediction. $\lambda_c$ represents the weight that controls how much the error on class $c$ contributes to the loss, $c = 1, ..., C$. Finally, the loss $\mathcal{L}_M$ of the whole MUL module is derived by averaging the loss values of all the experts:
\begin{equation}
\label{loss_MUL}
  \mathcal{L}_M = \frac{1}{N}\sum^{N}_{j=1} \mathcal{L}_e^j.
\end{equation}

We use the variance of classification loss $\mathcal{L}_e^j$ with regards to the average loss $\mathcal{L}_M$ to quantify the uncertainty, denoted by $\sigma$, which is given as:
\begin{equation}
\label{sigma}
  \sigma = \frac{1}{N}\sum^{N}_{j=1} (\mathcal{L}_M-\mathcal{L}_e^j)^2.
\end{equation}
The proposed MUL module improves the diagnostic accuracy as the final prediction combines the results from multiple experts, and also mitigates the negative effects caused by the noise in the data by introducing the dropout layers. Moreover, the experiments have revealed that the more experts in MUL module the faster the system converges during training.

\subsection{Training}
The whole architecture of RCoNet$^k_s$ is presented in Fig.~\ref{framework}, where the CXR images are first processed by a stack of deformable convolution layers, then transformed to high-order moment latent features by the MHMF module, which are then fed to the MUL module to generate final diagnoses. The loss used to optimize RCoNet$^k_s$ is given as follows
\begin{equation}
\label{loss_total}
  \mathcal{L}_{total} = \mathcal{L}_M-\alpha\mathcal{L}_I,
\end{equation}
where $\mathcal{L}_M$ is the prediction loss given by Eq.~\eqref{loss_MUL} , and $\mathcal{L}_I$ denotes the mutual information between the input $X$ and the latent representation $E_\psi(X)$ estimated by either Eq.~\eqref{loss_jsd} or Eq.~\eqref{nce}. $\alpha$ is a positive hyper-parameter that governs how much $\mathcal{L}_M$ and $\mathcal{L}_I$ contribute to the total loss. 
During training, the trainable parameters of the whole systems are updated iteratively to minimize $\mathcal{L}_{total}$, which is to jointly minimize the prediction loss $\mathcal{L}_M$ thus to improve the accuracy, and maximize the mutual information $\mathcal{L}_I$.

%% file: results.tex
\section{Experiments and Results}
\label{sec_experiment}
\subsection{{Dataset}}
\label{subsec_dataset}
We use a public chest X-ray dataset, referred to as \textit{COVIDx}, to evaluate the proposed model,  which is published by the authors of COVID-Net \cite{wang2020covid}. This dataset contains a total of 13975 CXR images from 13870 patients of 3 classes: (a) normal (no infections); (b) pneumonia (non-COVID-19 pneumonia); (c) COVID-19. It contains samples from five open source available data repositories {\em https://github.com/lindawangg/COVID-Net/blob/master/docs/COVIDx.md}. Three random CXR samples of these three classes are shown in Fig.~\ref{sample}. To reduce the negative effect caused by extremely unbalanced training samples, i.e., very limited number of COVID-19 positive cases compared to the other two categories, we further include other open-source CXR datasets from {\em https://www.kaggle.com/c/rsna-pneumonia-detection-challenge/data}. Following~\cite{wang2020covid,ahmed2020reconet}, the dataset is finally divided into 13624 training and 1510 test samples. The numbers of samples from different categories used for training and testing are summarized in Table~\ref{Table_details}. Moreover, we also adopted various data augmentation techniques to generate more COVID-19 training samples, such as flipping, translation, rotation using random five different angles, to tackle the data imbalance issue such that the proposed model can learn an effective mechanism of detecting COVID-19.  
%%%%%%%%%%%%%%% begin table 1 %%%%%%%%%%%%%%%%%%%%%%%%
\begin{table}[t]
    \caption{Details of patient data used for training and testing}
    \label{Table_details}
    \begin{center}
    \setlength{\tabcolsep}{3mm}{
        \begin{tabular}{|c|c|c|c|c|}
            \hline
             \multirow{2}{*}{Data} & \multicolumn{3}{c|}{Number of Patients Per Class} & \multirow{2}{*}{Total Patients}  \\ \cline{2-4}
             & Normal  & Pneumonia & COVID-19 & \\ \hline
             Train & 7966 & 5451 & 207 & 13624 \\ \hline
             Test  & 885 & 594 & 31 & 1510 \\ \hline
        \end{tabular}}
    \end{center}
\end{table}
%%%%%%%%%%%%%%% end table 1 %%%%%%%%%%%%%%%%%%%%%%%%
\subsection{{Evaluation Metrics}}
\label{subsec_eva_metrics}
In our experiments, we use the following six metrics to evaluate the COVID-19 detection performance of different approaches: 
\begin{itemize}
    \item Accuracy ($ACC$): $ACC$ calculates the proportion of images that are correctly identified. $ACC=\frac{TP+TN}{TP+TN+FP+FN}$.
    \item Sensitivity ($SEN$): $SEN$ is the ratio of the  positive cases that have been correctly detected to all the positive cases. $SEN=\frac{TP}{TP+FN}$.
    \item Specificity ($SPE$): $SPE$ is the ratio of the  negative cases that have been correctly classified to all the negative cases. $SPE=\frac{TN}{TN+FP}$.
    \item Balance ($BAC$): $BAC$ is  the mean value of $SEN$ and $SPE$. $BAC=\frac{SEN+SPE}{2}$.
    \item Positive Predictive Value ($PPV$): $PPV$ is the ratio of correctly detected positive cases to all cases that are detected to be positive. $PPV=\frac{TP}{TP+FP}$. 
    \item F1-score ($F1$): $F1$ uses a combination of accuracy and sensitivity to calculate a balanced average result. $F1=\frac{2\times ACC\times SEN}{ACC+SEN}$.
\end{itemize}
$TN$, $TP$, $FN$ and $FP$ represent the total number of true negatives, true positives, false negatives, and false positives, respectively.

\subsection{{Compared Methods}}
\label{subsec_com_methods}
We compare the proposed RCoNet$^k_s$ with the following five existing deep learning methods for COVID-19 detection:
\begin{itemize}
    \item PbCNN~\cite{oh2020deep}: A patch-based convolutional neural network with a relatively small number of trainable parameters.
    \item COVID-Net~\cite{wang2020covid}: A tailored deep convolutional neural network that uses a projection-expansion-projection design pattern.
    \item DenseNet-121~\cite{huang2017densely}: A densely connected convolutional network that connects each layer to every other layer in a feed-forward fashion.
    \item CoroNet~\cite{khan2020coronet}: A deep convolutional neural network model based on Xception architecture pre-trained on ImageNet dataset.
    \item ReCoNet~\cite{ahmed2020reconet}: A residual image-based COVID-19 detection network that exploits a CNN-based multi-level preprocessing filter block and a multi-task learning loss.
\end{itemize}

%%%%%%%%%%%%%%% begin table 2 %%%%%%%%%%%%%%%%%%%%%%%%
\begin{table}[t]
    \caption{Details of 10\% noisy patient data used for training.}
    \label{Table_noise_matrix}
    \begin{center}
    \setlength{\tabcolsep}{3mm}{
        \begin{tabular}{|c|c|c|c|}
            \hline
             Training Date & Clean  & Noise     & Total   \\ \hline
             Normal        & 7170   & 796 (Peumonia+COVID-19) & 7966    \\ \hline
             Pneumonia     & 4906   & 545 (COVID-19+Normal) & 5451    \\ \hline
             COVID-19      & 187    & 20  (Peumonia+Normal) & 207     \\ \hline
        \end{tabular}
        }
    \end{center}
\end{table}
%%%%%%%%%%%%%% end table 2 %%%%%%%%%%%%%%%%%%%%%%%%

%%%%%%%%%%%%%%% begin table 3 %%%%%%%%%%%%%%%%%%%%%%%%
\begin{table*}[htp]
  \centering
  \caption{Performance comparison of different approaches for COVID-19 detection on the COVIDx dataset}
  \label{Table_all_metrics}
%   \resizebox{\textwidth}{!}{
\setlength{\tabcolsep}{2.6mm}{
  \begin{tabular}{c|cccccccc}
\hline
Method&ACC ($\%$)&SEN ($\%$)&SPE ($\%$)&BAC ($\%$)&PPV ($\%$)&F1 ($\%$)&Param (M)&FLOPs (G) \\
\hline
PbCNN~\cite{oh2020deep}&88.90$\pm$1.63 &85.90$\pm$1.69 &96.40$\pm$2.10 &91.15$\pm$1.31 &88.65$\pm$1.52 &87.37$\pm$2.14  &11.60 &-\\

COVID-Net~\cite{wang2020covid}&95.10$\pm$1.34 &91.37$\pm$1.37 &95.76$\pm$2.04 &93.57$\pm$0.89 &94.73$\pm$0.97 &93.20$\pm$0.85 &117.4 &15.10\\

DenseNet-121~\cite{huang2017densely}&97.40$\pm$1.67 &96.08$\pm$0.88 &97.23$\pm$1.01 &96.66$\pm$1.21 &96.05$\pm$1.00 &96.74$\pm$1.04 &7.61 &bf{5.59} \\

CoroNet~\cite{khan2020coronet}&95.00$\pm$1.58 &96.90$\pm$1.57 &97.50$\pm$1.93 &97.20$\pm$1.07 &95.00$\pm$1.03 &95.60$\pm$0.95 &33.00 &- \\

ReCoNet~\cite{ahmed2020reconet}&97.48$\pm$1.05 &97.39$\pm$1.67 &97.53$\pm$1.28 &97.46$\pm$0.87 &97.17$\pm$0.76 &97.43$\pm$0.59 &\bf{2.52} &7.68 \\

\hline
RCoNet$^1_4$&96.12$\pm$0.33 & 95.71$\pm$0.41 & 96.38$\pm$0.29 & 96.05$\pm$0.20 &95.86$\pm$0.62  &95.91$\pm$0.56 &.73 &7.61 \\

RCoNet$^2_4$&96.78$\pm$0.57 & 96.48$\pm$0.69 & 96.91$\pm$0.74 & 96.70$\pm$0.34 &96.94$\pm$0.53  &96.63$\pm$0.58 &6.74 &7.70 \\

RCoNet$^3_4$&97.46$\pm$0.43 & 97.25$\pm$0.79 & 97.62$\pm$0.40 & 97.44$\pm$0.82 &97.59$\pm$0.91  &97.35$\pm$0.38 &6.75 &7.79 \\

RCoNet$^4_4$&\bf{97.89$\pm$0.53} & 97.33$\pm$0.45 & \bf{98.24$\pm$0.39} & \bf{97.79$\pm$0.62} &\bf{97.93$\pm$0.74}  &97.61$\pm$0.48 &6.77 &7.91 \\

RCoNet$^5_4$&97.50$\pm$0.62 &\bf{97.76$\pm$0.87} &97.18$\pm$0.63 &97.47$\pm$0.73 &97.10$\pm$0.91 &\bf{97.63$\pm$0.71} &6.77 &8.00 \\
\hline
    \end{tabular}
    }
\end{table*}
%%%%%%%%%%%%%%% end table 3 %%%%%%%%%%%%%%%%%%%%%%%%
%%%%%%%%%%%%%%% begin figure 4 %%%%%%%%%%%%%%%%%%%%%%%%
\begin{figure*}[tp]
\centering
\subfigure[Clean]{
\begin{minipage}[t]{0.24\linewidth}
% \label{con:a}
\centering
\includegraphics[width=1.7in]{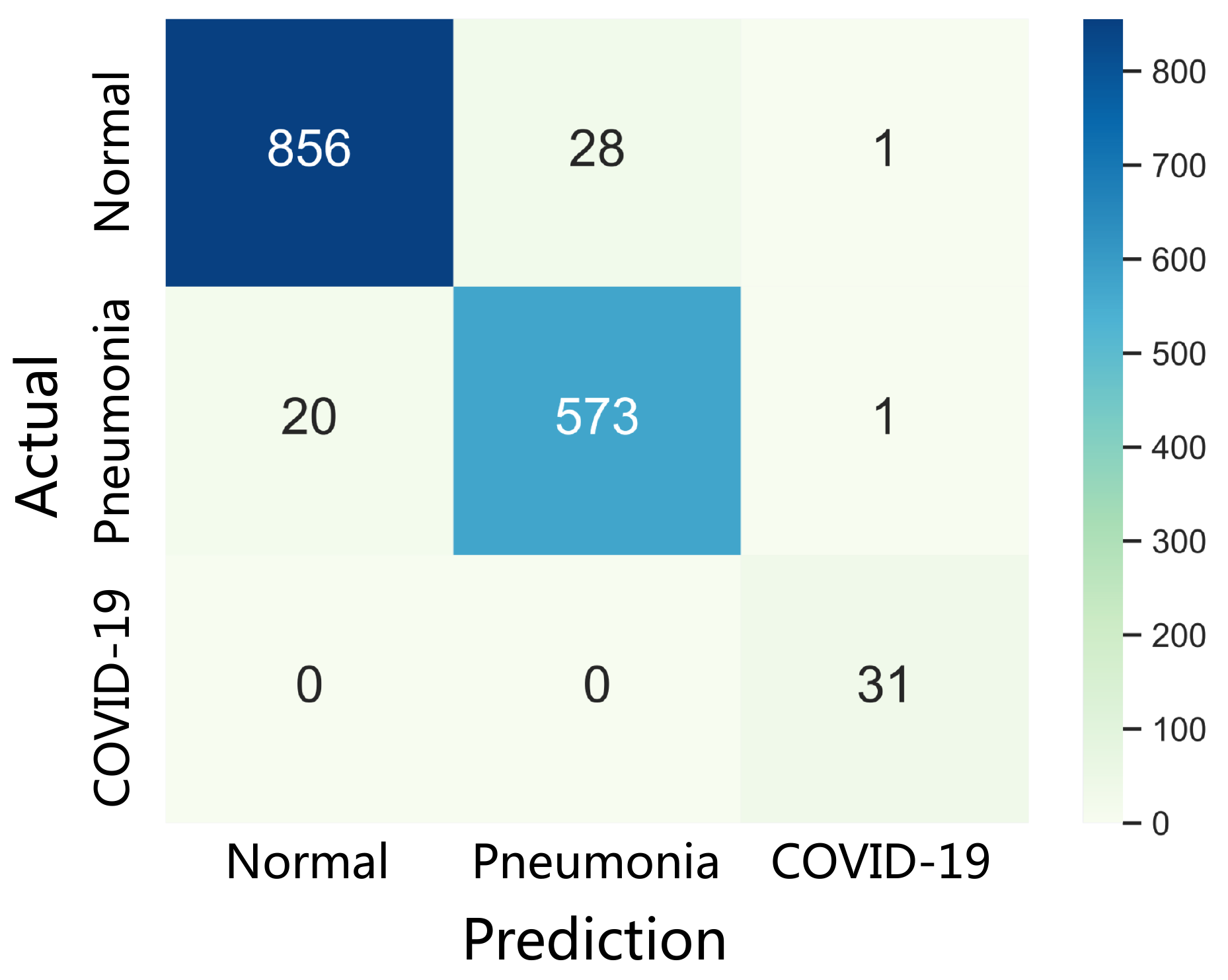}
\end{minipage}%
}%
\subfigure[$10\%$ Noise]{
\begin{minipage}[t]{0.24\linewidth}
% \label{con:b}
\centering
\includegraphics[width=1.7in]{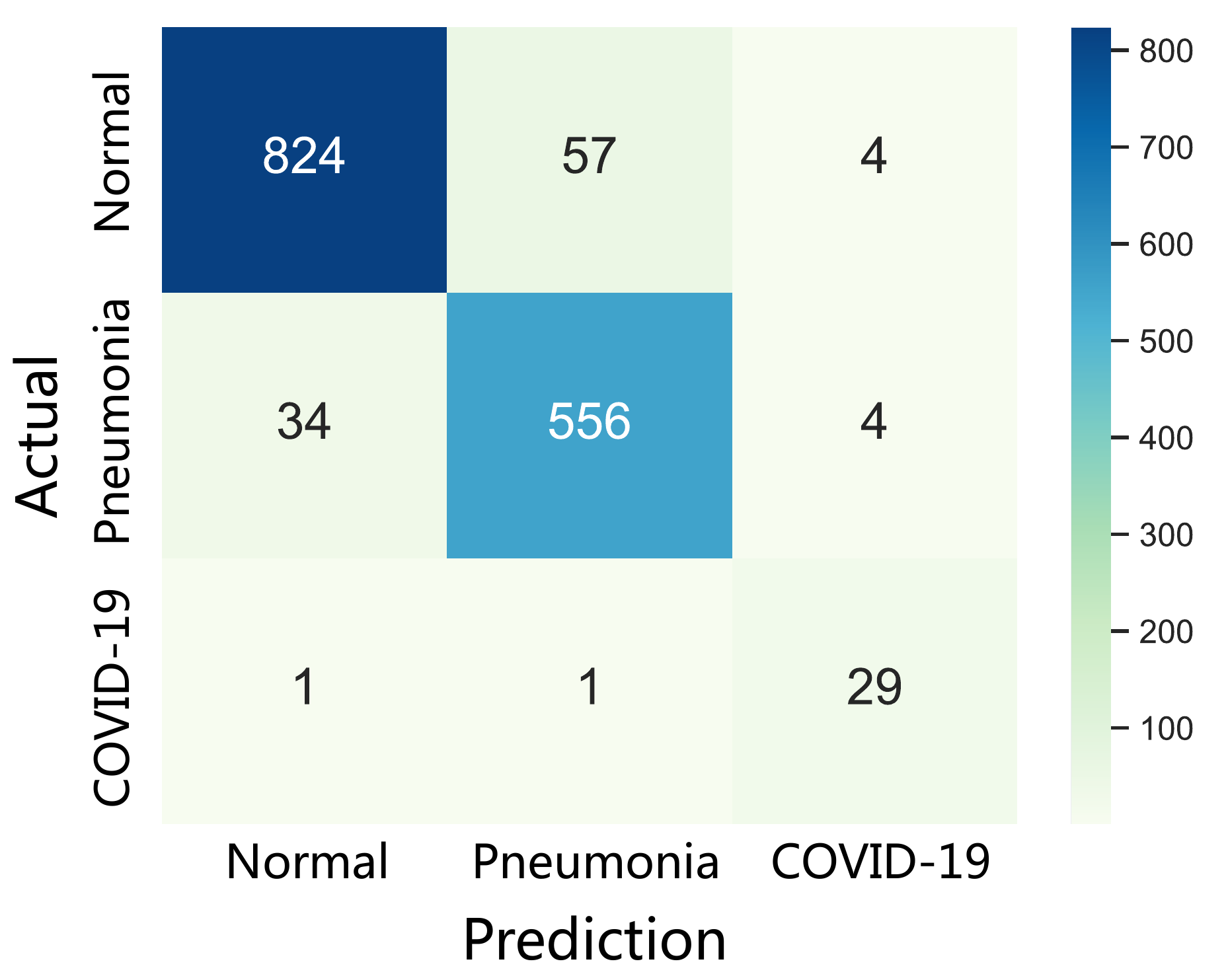}
\end{minipage}%
}%
\subfigure[$20\%$ Noise]{
\begin{minipage}[t]{0.24\linewidth}
% \label{con:c}
\centering
\includegraphics[width=1.7in]{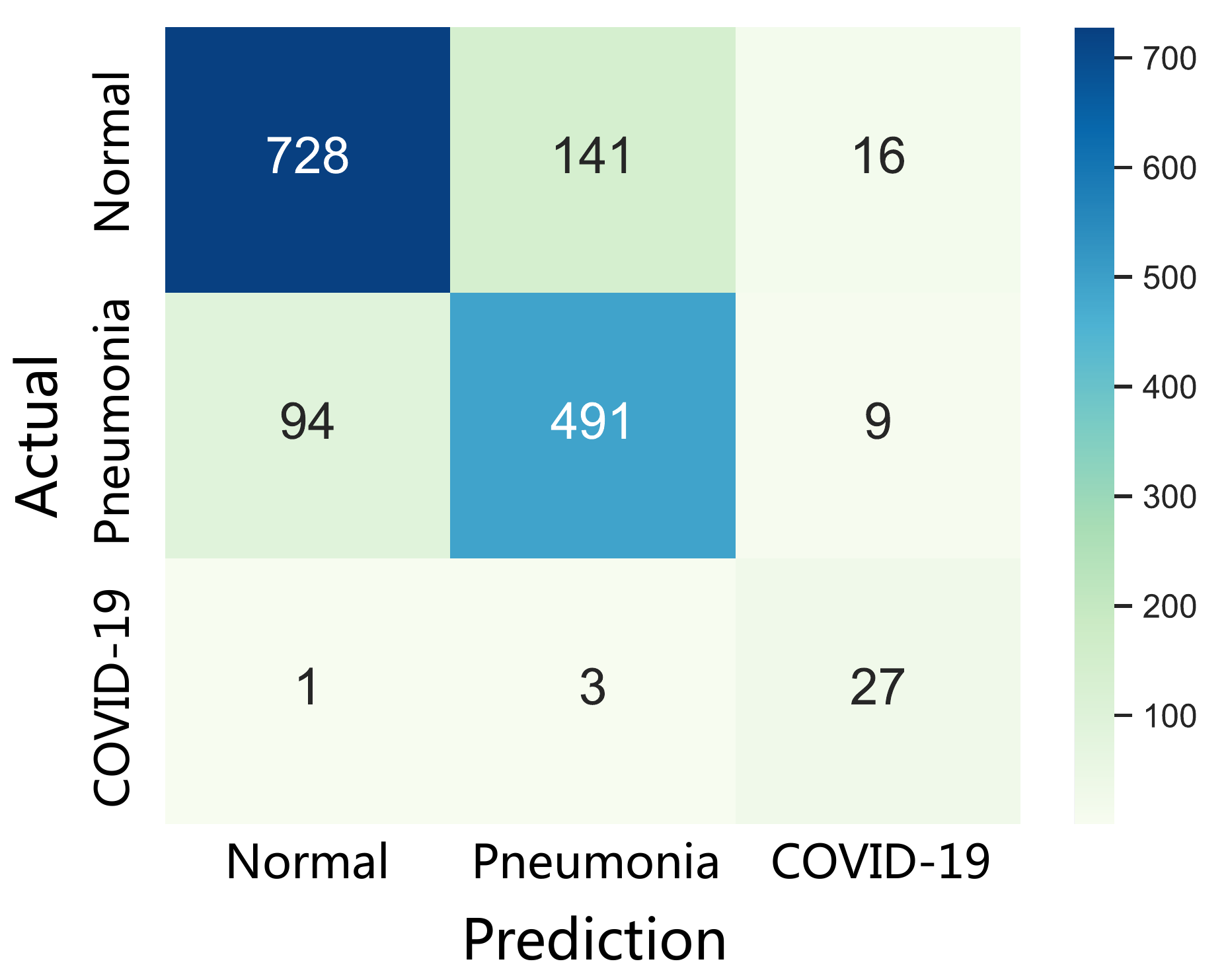}
\end{minipage}
}%
\subfigure[$30\%$ Noise]{
\begin{minipage}[t]{0.24\linewidth}
% \label{con:d}
\centering
\includegraphics[width=1.7in]{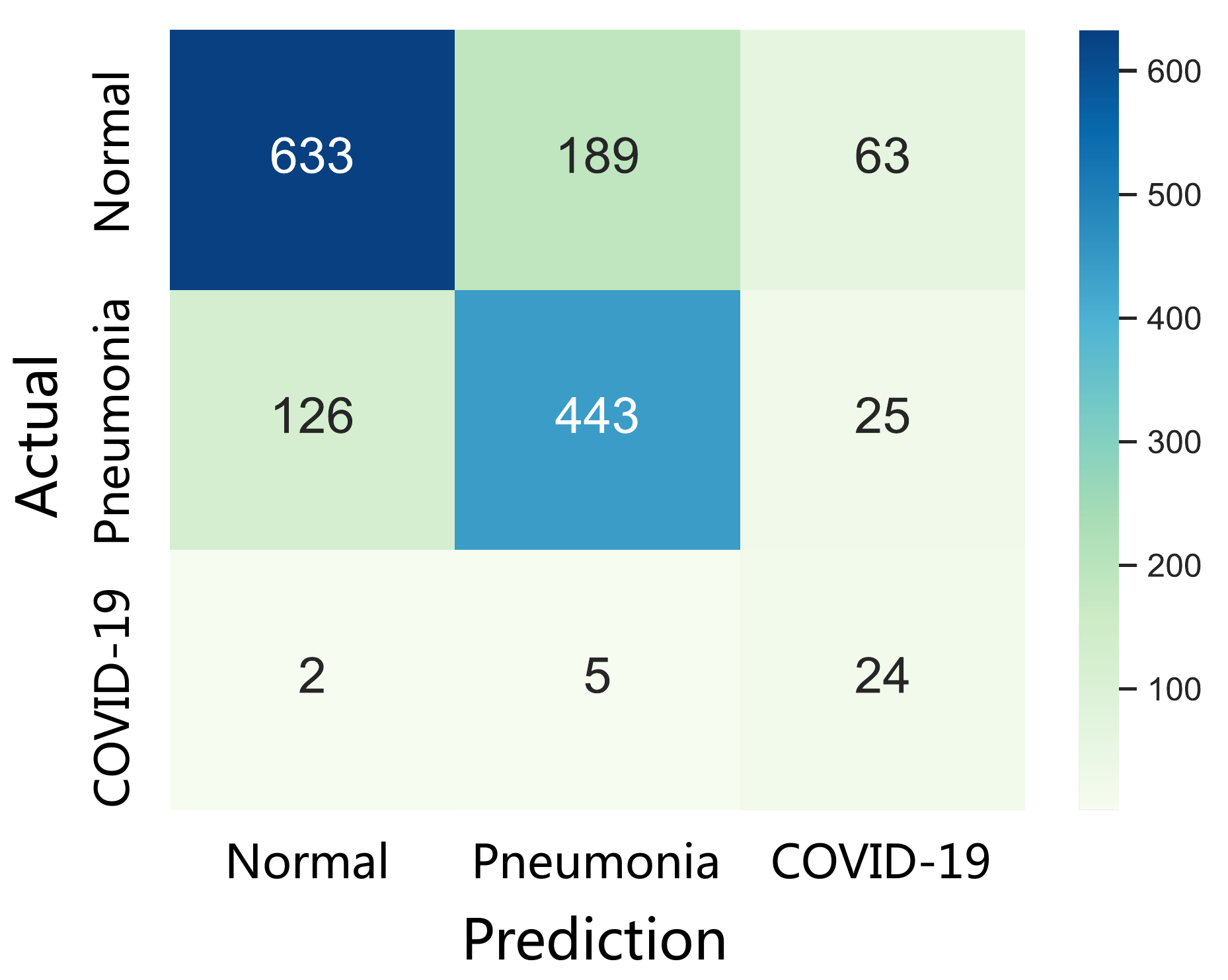}
\end{minipage}
}%
\centering
\caption{Confusion matrices of the proposed RCoNet$^k_s$ trained on noisy dataset with different percentages of noisy samples.}
\label{fig_con}
\end{figure*}
%%%%%%%%%%%%%%% end figure 4 %%%%%%%%%%%%%%%%%%%%%%%%

\subsection{{Implementation}}
\label{subsec_implementation}
We implement our RCoNet$^k_s$ using the PyTorch library and apply ResNeXt~\cite{xie2017aggregated} as the backbone network. We train the model with the Adam optimizer with an initial learning rate of $2\times10^{-4}$ and a weight decay factor of $1\times10^{-4}$. All the experiments are run on an NVIDIA GeForce GTX 1080Ti GPU. We set the batch size to be 8, and resize all images to $224\times224$ pixels. The hyperparameter $\alpha$ in the loss function given in Eq.~\eqref{loss_total} is set to be within the range of $[0,0.4]$. The drops rate of each dropout layer in the MUL module is randomly chosen from $\{0.1, 0.3, 0.5\}$. The loss weight $\lambda_c$ for each category, which is used to calculate the weighted sum of the loss as given in Eq.~\eqref{loss_weight}, is set to be $1$, $1$, and $20$ for the normal, pneumonia, COVID-19 samples, respectively, corresponding to the number of training samples in each. We adopt 5-fold cross-validation training that we randomly divide the training sets into five equal-size subsets and train the model five times that using different four subsets for training, and the remaining one for validation each time. We also evaluate our proposed model with different number of order moments for the MHMF module $k$, and different number of experts $s$.

To evaluate the performance of the proposed model with the presence of label noise, we derive a noisy dataset from the given dataset in the following way: we randomly select a given percentage of training samples in each category, and assign wrong labels to these sample. In particular, to ensure that the fake COVID-19 samples are less than the real ones, we assign the COVID-19 labels to selected normal and pneumonia samples in a way the the number of normal and pneumonia samples assigned with COVID-19 label equals to the number of COVID-19 samples assigned with either normal and pneumonia label. We show a realization of the derived noisy dataset when the percentage of fake samples is set to be 10$\%$ in Table~\ref{Table_noise_matrix}. 
%%%%%%%%%%%%%%% begin table 4 %%%%%%%%%%%%%%%%%%%%%%%%
\begin{table}[t]
  \centering
  \caption{Performance comparison of different approaches on COVIDx dataset with noisy samples}
  \label{Table_noise}
  \begin{tabular}{c|c|ccc}
\hline
Noise&Method&ACC ($\%$)&SEN ($\%$)&SPE ($\%$)\\
\hline
\multirow{7}{*}{10$\%$}
&PbCNN~\cite{oh2020deep}&83.22 &81.98 &89.01\\

&COVID-Net~\cite{wang2020covid}&91.03 &87.94 &90.62 \\

&DenseNet-121~\cite{huang2017densely}&91.97 &87.94 &92.17\\

&CoroNet~\cite{khan2020coronet}&89.45 &88.74 &90.06\\

&ReCoNet~\cite{ahmed2020reconet}&91.63 & 90.82 &91.16\\ \cline{2-5}

&RCoNet$^3_4$&92.78 &92.21 &\bf{93.51} \\

&RCoNet$^4_4$&\bf{92.98} &\bf{93.39} &93.12 \\

&RCoNet$^5_4$&92.01 &91.41 &92.76 \\\hline
% \cline{2-5}
\multirow{7}{*}{20$\%$}
&PbCNN~\cite{oh2020deep}&78.42 &75.90 &80.29\\

&COVID-Net~\cite{wang2020covid}&82.51 &82.77 &81.95\\

&DenseNet-121~\cite{huang2017densely}&82.16 &81.01 &82.21\\

&CoroNet~\cite{khan2020coronet}&82.33 &81.10 &81.89\\

&ReCoNet~\cite{ahmed2020reconet}&83.26 &82.72 &83.17\\ \cline{2-5}

&RCoNet$^3_4$&84.18 &84.56 &85.79 \\

&RCoNet$^4_4$&84.30 &\bf{84.01} &\bf{85.99} \\

&RCoNet$^5_4$&\bf{84.34} &83.96 &85.21 \\\hline
% \cline{2-5}
\multirow{7}{*}{30$\%$}
&PbCNN~\cite{oh2020deep}&67.76 &66.47 &70.61\\

&COVID-Net~\cite{wang2020covid}&71.98 &70.13 &71.55\\

&DenseNet-121~\cite{huang2017densely}&72.74 &72.36 &72.96\\

&CoroNet~\cite{khan2020coronet}&71.87 &72.02 &71.54\\

&ReCoNet~\cite{ahmed2020reconet}&73.26 &72.53 &73.11\\ \cline{2-5}

&RCoNet$^3_4$&74.56 &74.20 &75.54 \\

&RCoNet$^4_4$&74.69 &\bf{74.51} &\bf{76.94} \\

&RCoNet$^5_4$&\bf{74.88} &74.37 &75.21 \\
\hline
    \end{tabular}
\end{table}
%%%%%%%%%%%%%%% end table 4 %%%%%%%%%%%%%%%%%%%%%%%%

\subsection{{Results and Discussions}}
{\bf Performance on Clean Data}:
The numerical results on the clean dataset without any artificial noise added are shown in Table~\ref{Table_all_metrics}. The results are presented in the form of $a\pm b$, where $a$ and $b$ denote the average and variance values of each metric on five independent experiments, respectively. We can see that RCoNet$^5_4$, i.e., the proposed model with $k=4$ levels of mixed moment features and $s=4$ experts, achieves notable performance improvement over the comparison methods in terms of most metrics considered, including ACC, SPE, BAC, PPV and F1 score. We note the performance of RCoNet$^k_s$ can be further improved with a different set of $k$ and $s$. For instance, RCoNet$^5_4$ achieves better SEN and F1 score than RCoNet$^4_4$. 
The higher ACC and F1 score validate that RCoNet$^k_s$ is able to obtain latent features, i.e., the mixed moment features of different levels of order, that maintains inter-class separability and intra-class compactness better than other models. Note that RCoNet$^5_4$ leads to a higher SEN than all other methods, which is particularly important to COVID-19 detection, since successfully detecting COVID-19 positive cases is the key to control the spread of this super contagious disease.
Moreover, it can be observed that RCoNet$^k_s$ has smaller variance compared to the others, which demonstrates the robustness and stability of our model. 
%%%%%%%%%%%%%%% begin figure 5 %%%%%%%%%%%%%%%%%%%%%%%%
\begin{figure*}[t]
\centering
\subfigure[baseline]{
\begin{minipage}[t]{0.24\linewidth}
\label{tSNE1:a}
\centering
\includegraphics[width=1.7in]{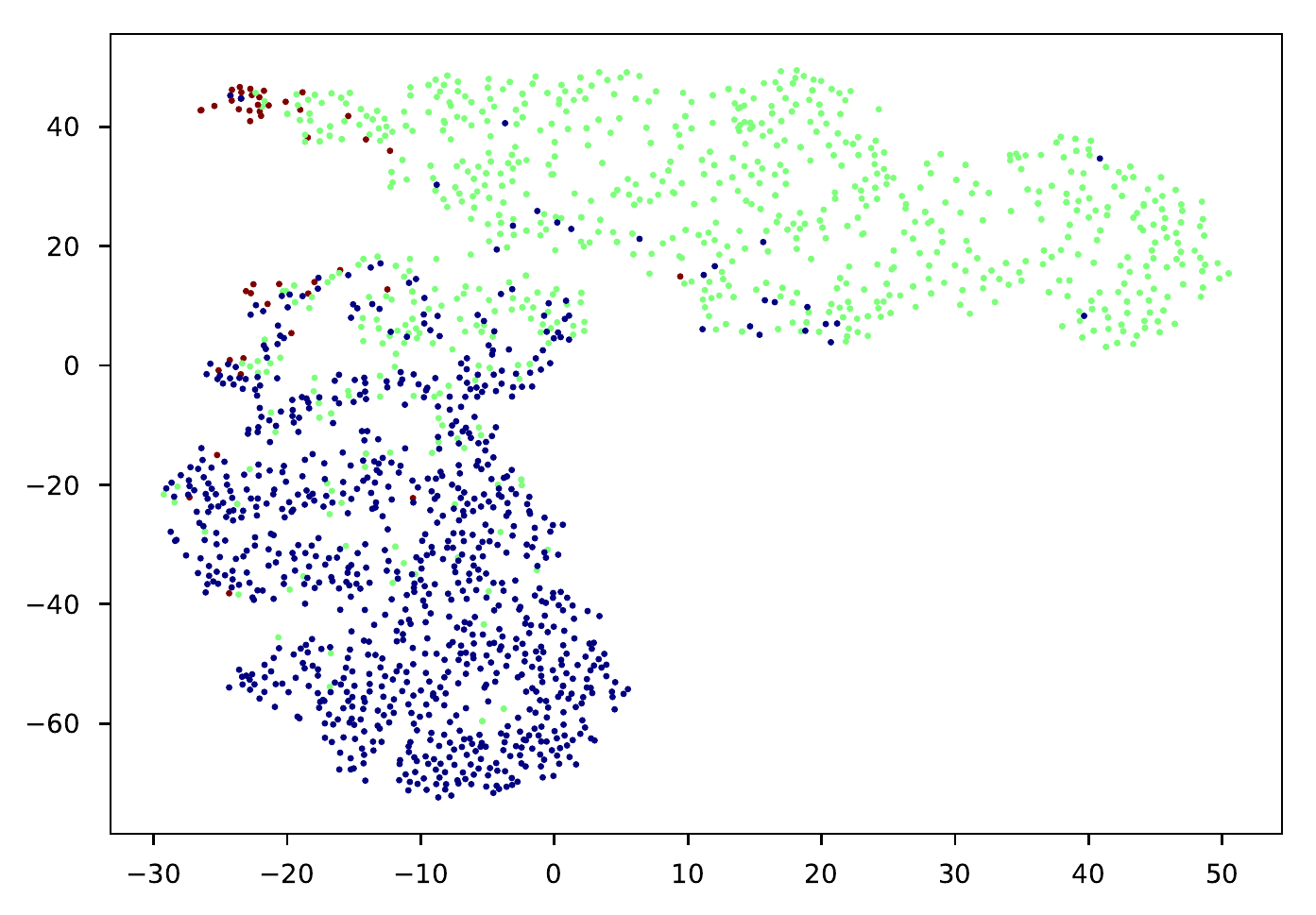}
\end{minipage}%
}%
\subfigure[RCoNet-D]{
\begin{minipage}[t]{0.24\linewidth}
\label{tSNE1:b}
\centering
\includegraphics[width=1.7in]{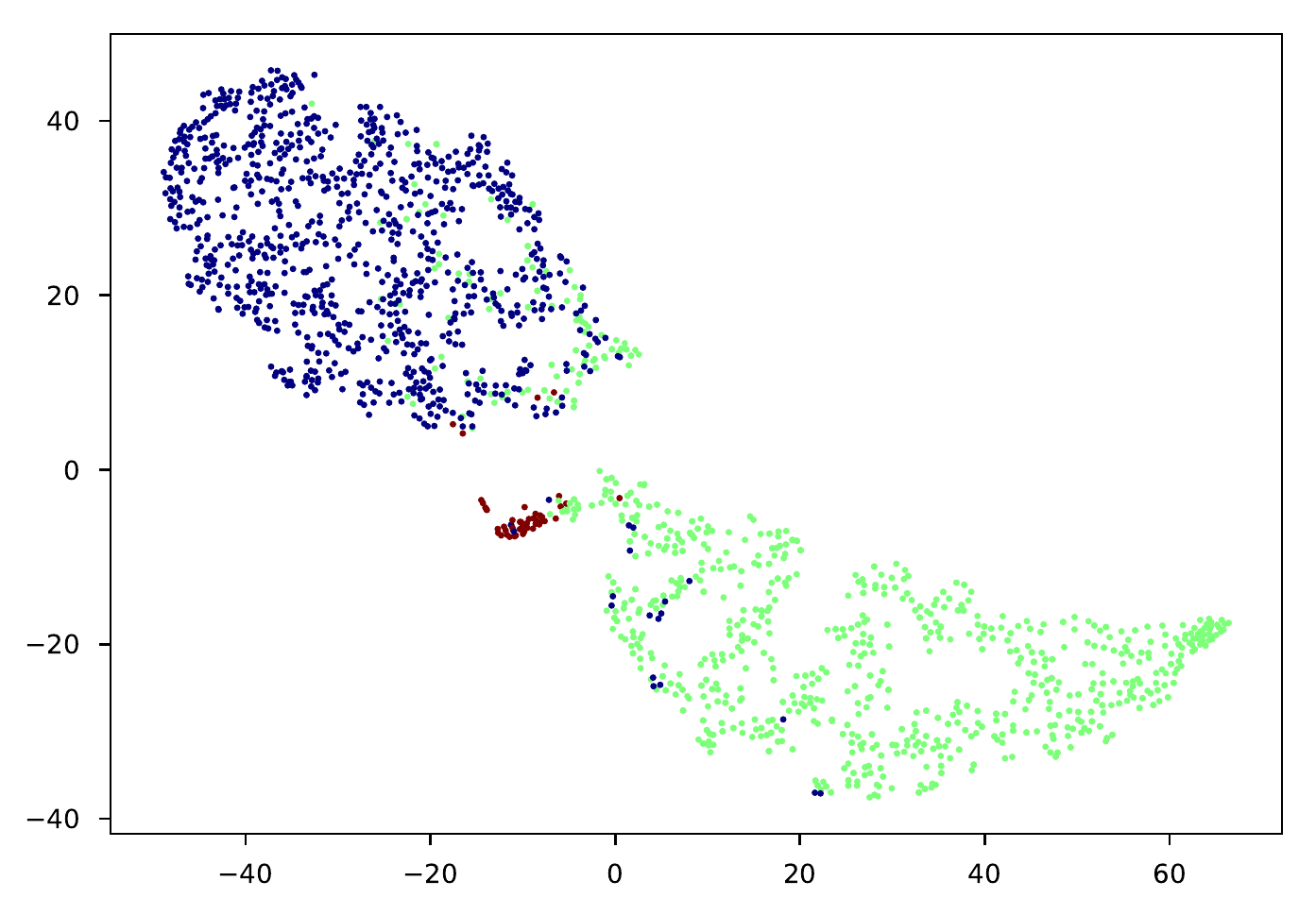}
\end{minipage}%
}%
\subfigure[RCoNet-M]{
\begin{minipage}[t]{0.24\linewidth}
\label{tSNE1:c}
\centering
\includegraphics[width=1.7in]{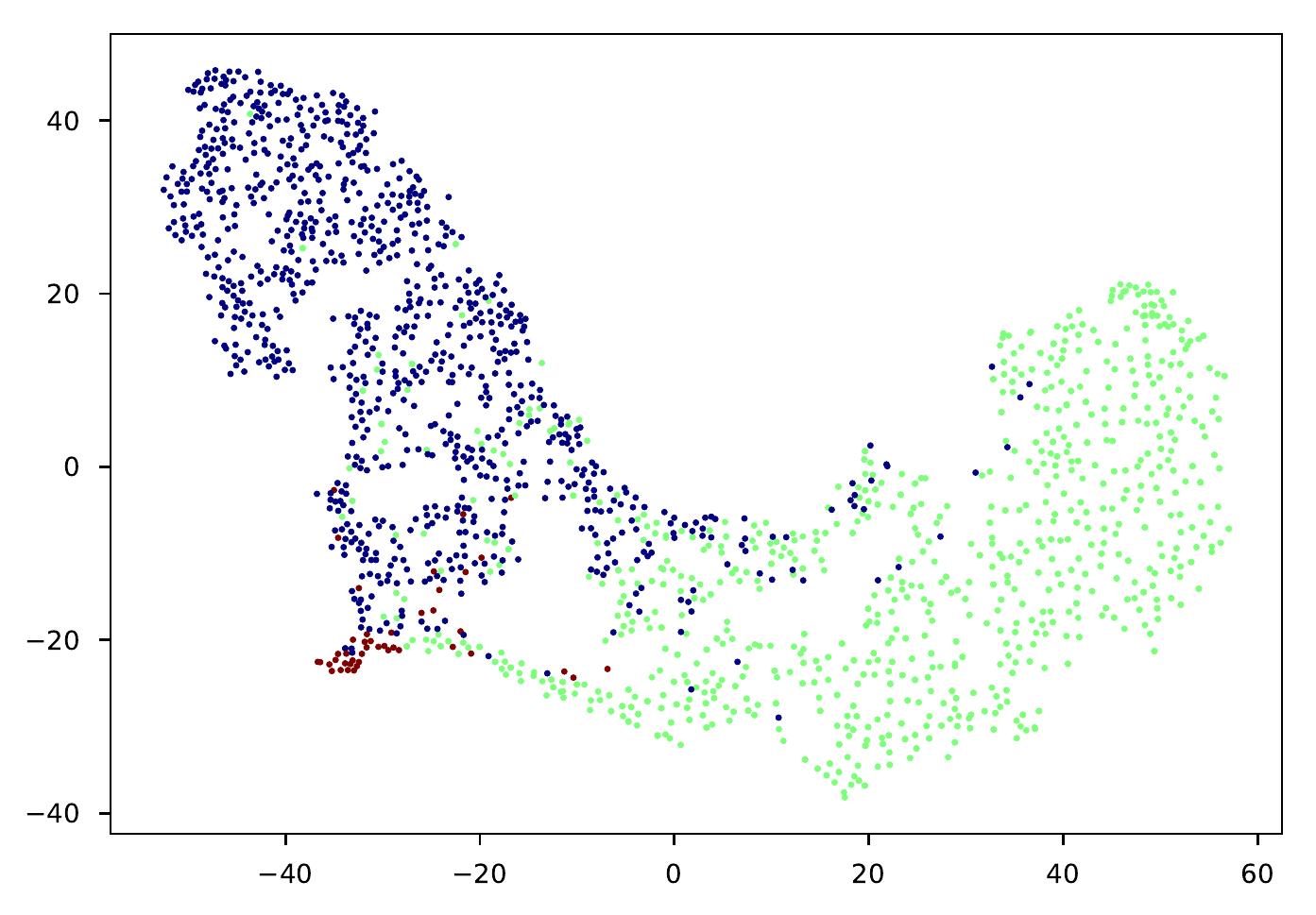}
\end{minipage}
}%
\subfigure[RCoNet-DM]{
\begin{minipage}[t]{0.24\linewidth}
\label{tSNE1:d}
\centering
\includegraphics[width=1.7in]{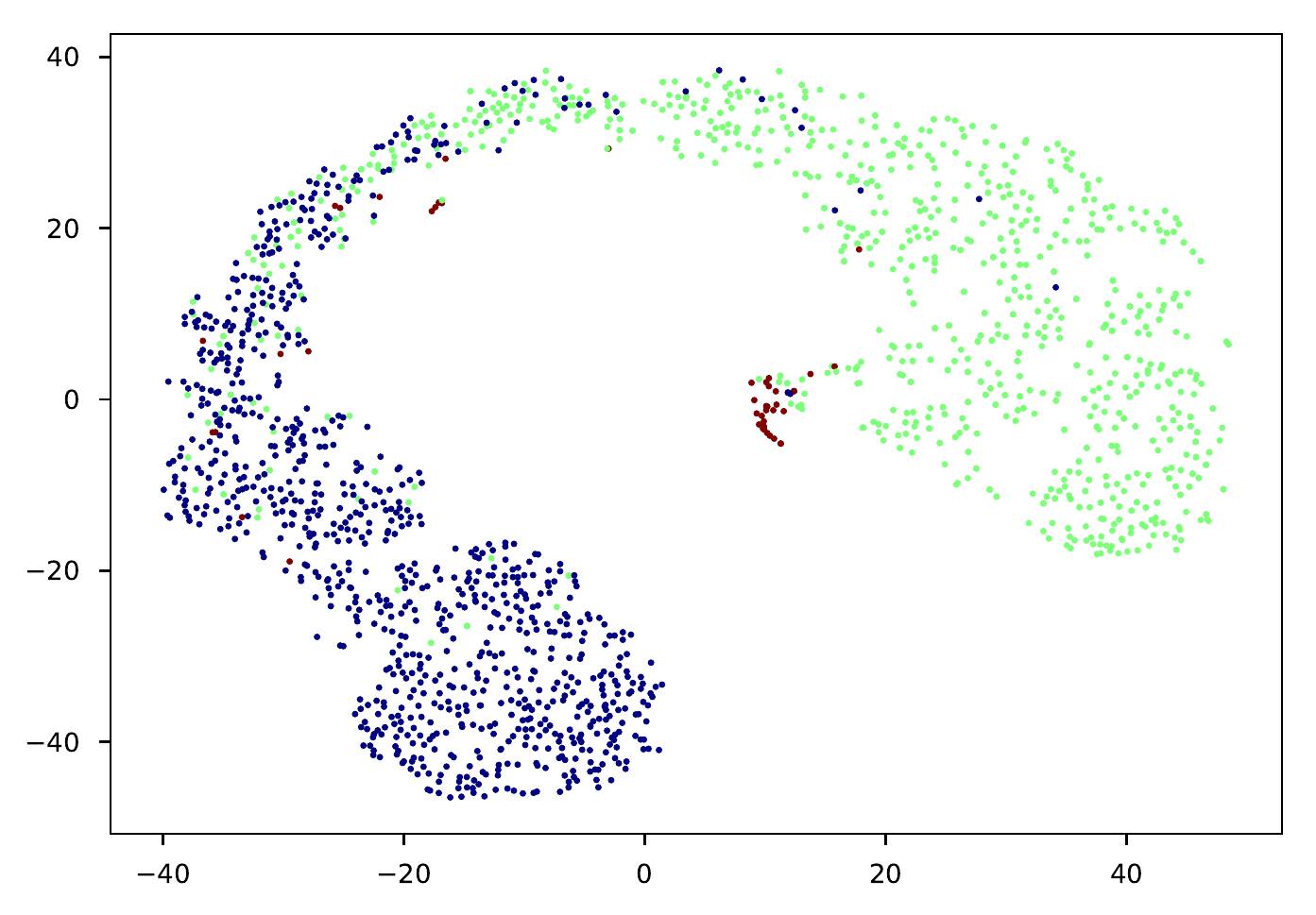}
\end{minipage}
}%

\subfigure[RCoNet$^2_4$]{
\begin{minipage}[t]{0.24\linewidth}
\label{tSNE1:e}
\centering
\includegraphics[width=1.7in]{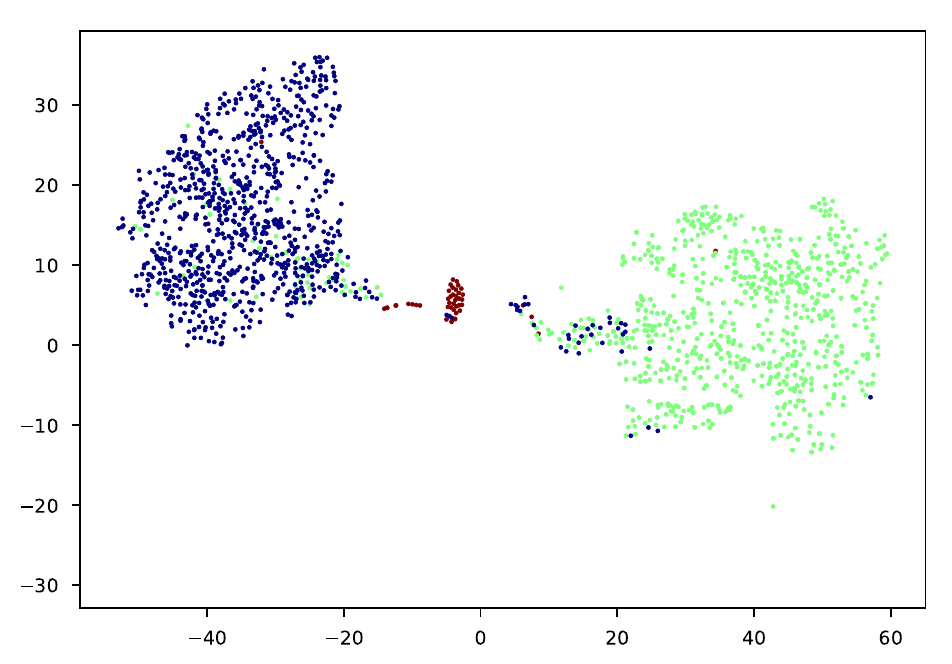}
\end{minipage}%
}%
\subfigure[RCoNet$^3_4$]{
\begin{minipage}[t]{0.24\linewidth}
\label{tSNE1:f}
\centering
\includegraphics[width=1.7in]{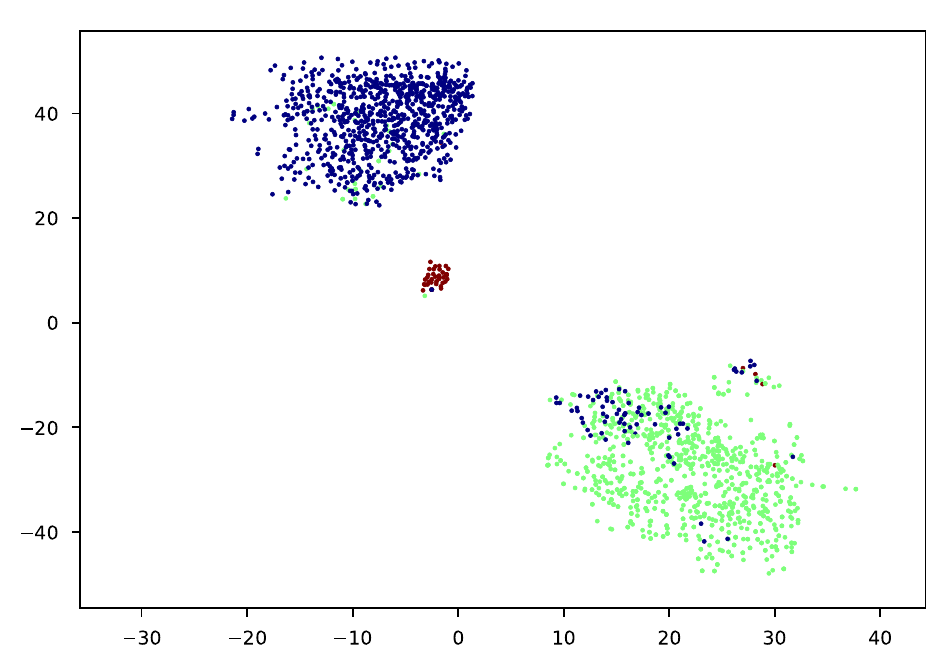}
\end{minipage}%
}%
\subfigure[RCoNet$^4_4$]{
\begin{minipage}[t]{0.24\linewidth}
\label{tSNE1:g}
\centering
\includegraphics[width=1.7in]{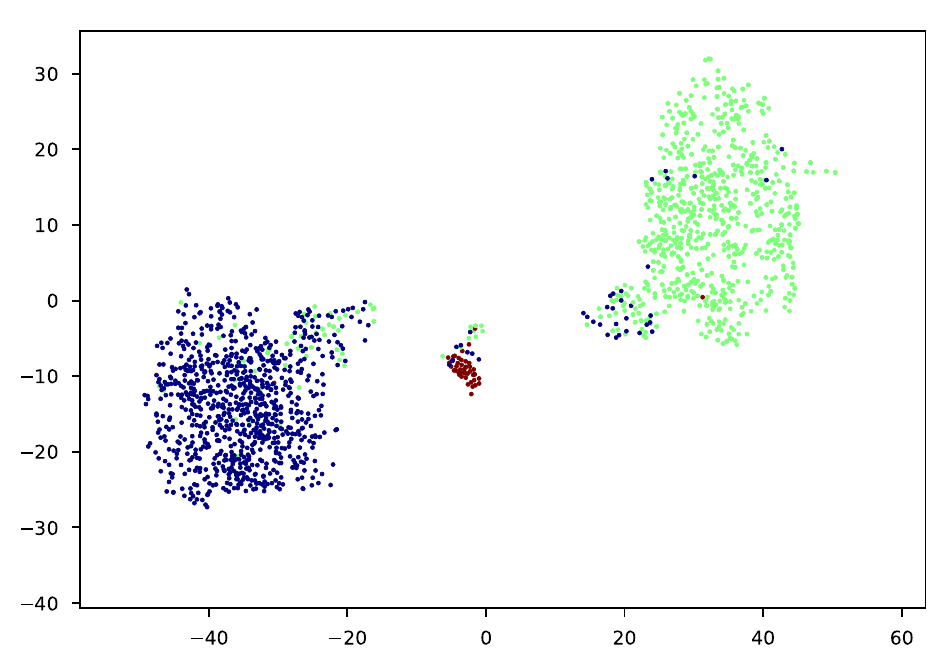}
\end{minipage}
}%
\subfigure[RCoNet$^5_4$]{
\begin{minipage}[t]{0.24\linewidth}
\label{tSNE1:h}
\centering
\includegraphics[width=1.7in]{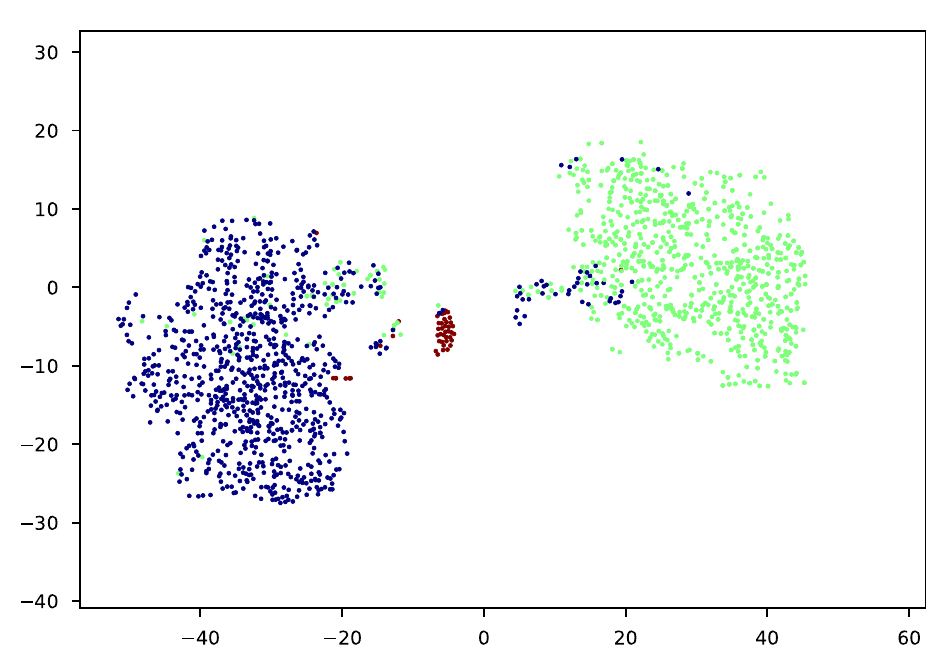}
\end{minipage}
}%
\centering
\caption{The t-SNE visualization of the latent features generated by different methods. Blue, green and red dots represent normal, pneumonia and COVID-19 samples, respectively.}
\label{fig_tSNE_all}
\end{figure*}
%%%%%%%%%%%%%%% end figure 5 %%%%%%%%%%%%%%%%%%%%%%%%

We also evaluate the complexity of the proposed model in terms of numbers of parameters and computational cost, i.e., Float-point operations (FLOPs), which is presented in Table~\ref{Table_all_metrics}. It can be observed that the proposed model has much fewer parameters than several existing methods, except ReCoNet. However, we note that the FLOPs of RCoNet$^k_s$ is quite close to that of ReCoNet, which means it takes a similar amount of time to diagnose COVID-19 from CXR images by these two model. We can also observe that the increase of $k$ and $s$, i.e., the number of mixed moment features and the number of experts in MUL, only causes a small, or even neglectable, amount of increase in the number of parameters and FLOPs as well, which suggests that we can improve the performance of the proposed model by optimizing $k$ and $s$, without the concern on the significant increase of the complexity. 
%%%%%%%%%%%%%%% begin figure 6 %%%%%%%%%%%%%%%%%%%%%%%%
\begin{figure}[t]
    \centering 
    \includegraphics[width = 8.8cm]{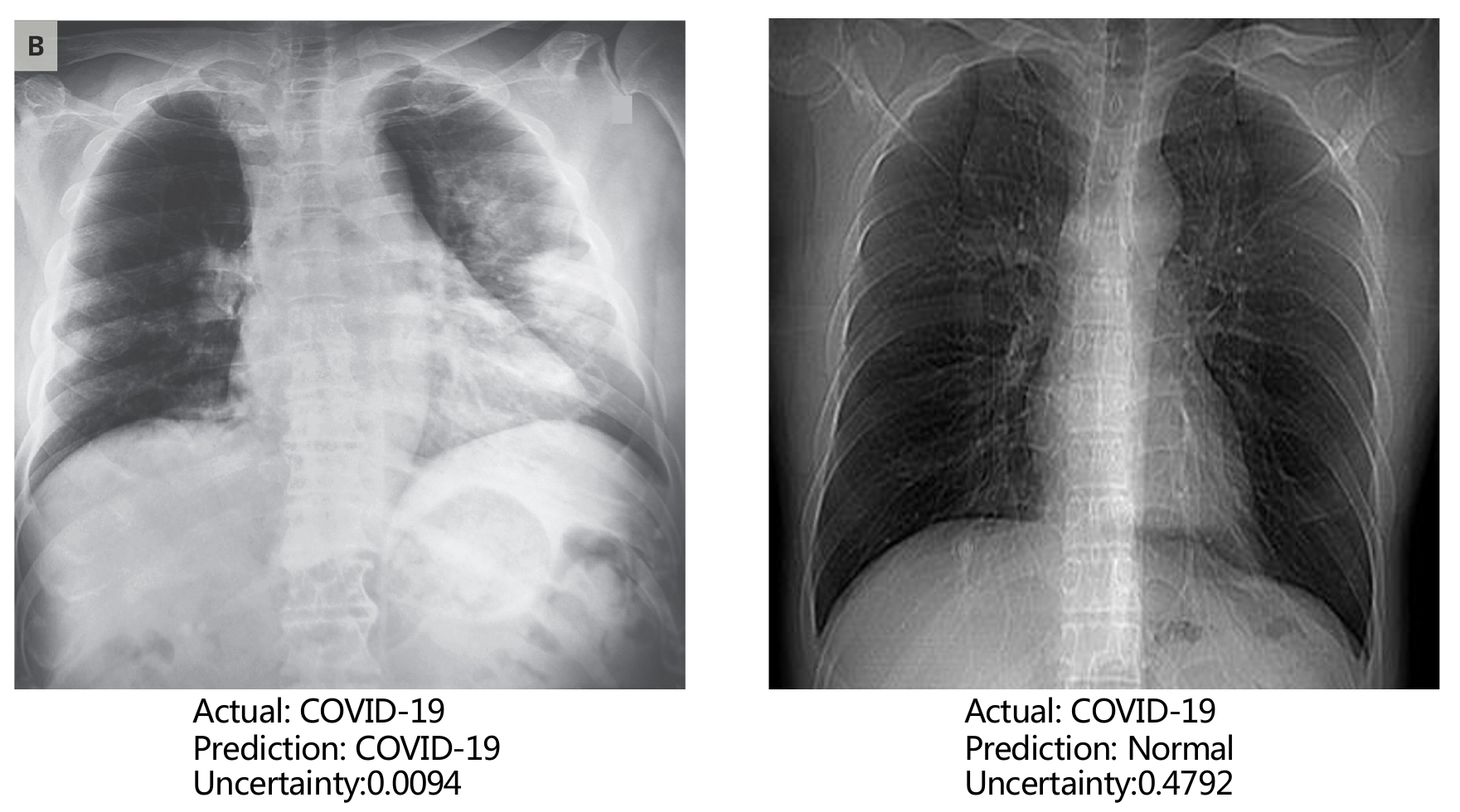}
    \caption{Example CXR samples with their predictions and the corresponding uncertainty levels by RCoNet$^4_4$.}
    \label{fig_uncertainty_sample}
\end{figure}
%%%%%%%%%%%%%%% end figure 6 %%%%%%%%%%%%%%%%%%%%%%%%

{\bf Performance on Noisy Data}:
We further compare the proposed model to the existing ones when there is noise present in the training dataset. We generate three noisy training datasets in the aforementioned way from the clean dataset with $10\%$, $20\%$ and $30\%$ samples with wrong labels, respectively. The results, which we take the averages from five independent experiments, are presented in Table~\ref{Table_noise}. It can be easily seen that the more fake samples we add the more it degrades the performance of all the methods. Note that the proposed RCoNet$^4_4$ still gets the state-of-the-art results in all considered cases with different percentages of noisy samples in the training dataset. Moreover, the performance gain over the existing methods slightly increases with the ratio of noisy samples, verifying that our model is more robust to the noise. Note that the extreme case of $30\%$ noisy samples leads to great performance degradation of all the models. In practice, the percentage of label noise is usually around $10\%$ to $20\%$. We present the confusion matrices in Fig.~\ref{fig_con} to summarize the prediction accuracy of different categories. We can observe that, although with very limited number of COVID-19, our model still maintains high accuracy of detecting COVID-19 cases, even with the presence of noisy samples.
%%%%%%%%%%%%%%% begin figure 7 %%%%%%%%%%%%%%%%%%%%%%%%
\begin{figure}[t]
    \centering 
    \includegraphics[width=8cm]{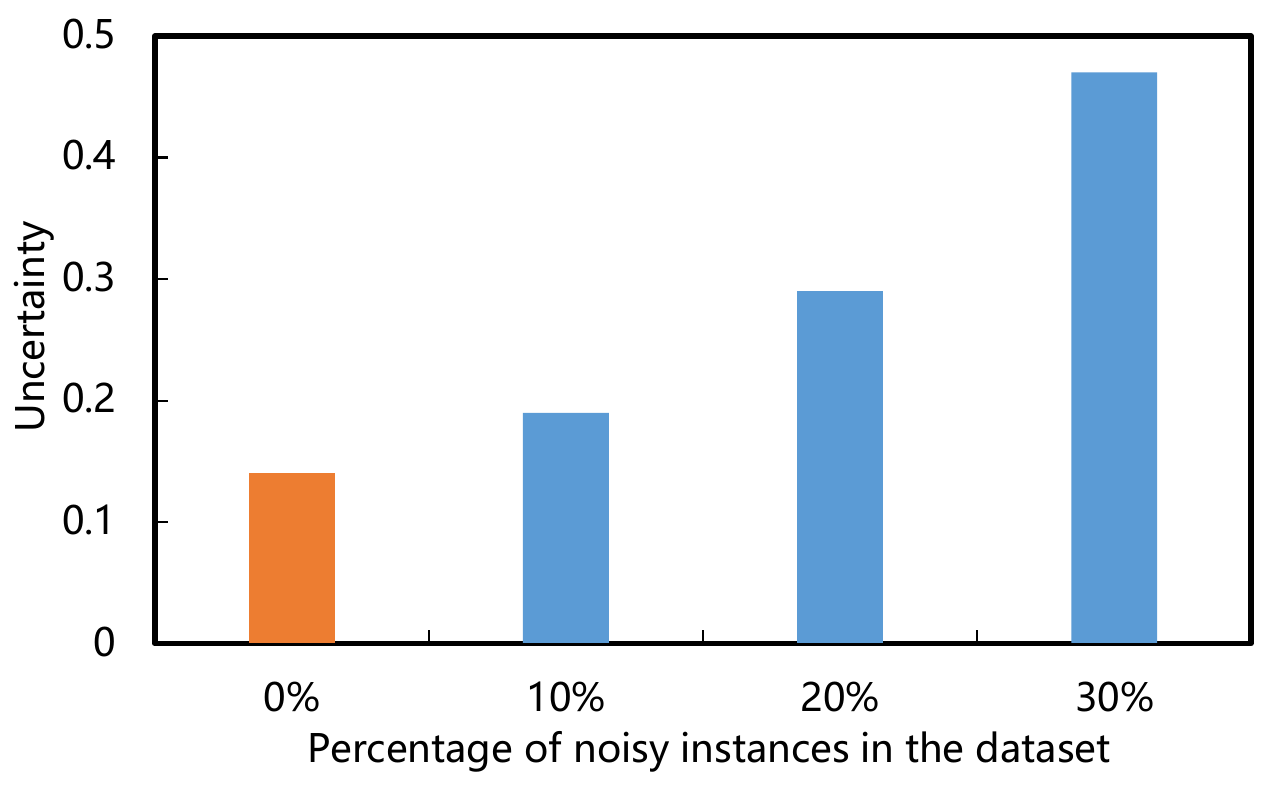}
    \caption{Comparison on uncertainty level of the predictions by RCoNet$^4_4$.}
    \label{fig_uncertainty}
\end{figure}
%%%%%%%%%%%%%%% end figure 7 %%%%%%%%%%%%%%%%%%%%%%%%

{\bf Uncertainty Estimation}:
One remarkable advantage of our model is the ability to quantify the uncertainty in the final prediction, which is significantly crucial for COVID-19 detection. This is done by obtaining the variance in the output of different experts in MUL as described in Section \ref{subsec_MUL}. The larger the variance is, the more different experts disagree with each other, and, hence, the more uncertain the model is about the final prediction. We present two CXR samples in Fig.~\ref{fig_uncertainty_sample}, including the predictions and the corresponding uncertainty level by RCoNet$^k_s$. We can see that the correctly classified CXR image has a low uncertainty level about its prediction, i.e., 0.0094, and the misclassified CXR sample with a high uncertainty level, i.e., 0.4792, suggests that an alternative way of diagnosis should be sought to correct this prediction. This greatly improves the reliability of the prediction by RCoNet$^k_s$, and reduces the chance of misdiagnosis. We also show in Fig.~\ref{fig_uncertainty} the average uncertainty levels of RCoNet$^k_s$ trained on clean and noisy datasets with different ratios of noisy samples. It can be observed that the uncertainty level increases almost linearly with the percentage of noisy samples in the dataset, which highlights the negative impact of noise on model training.

\subsection{{Analysis}}
We further numerically analyse the benefits of the three key modules of RCoNet$^k_s$, i.e., the DeIM, MHMF and MUL modules in this section.
%%%%%%%%%%%%%%% begin table 5 %%%%%%%%%%%%%%%%%%%%%%%%
\begin{table}[t]
  \centering
  \caption{Impact of the MHMF and MUL on the model performance.}
  \label{table_ablation}
  \setlength{\tabcolsep}{2.5mm}{
    \begin{tabular}{|c|c|c|c|c|c|c|c|}
    \hline
    RCoNet$^k_s$&s=1&s=2&s=3&s=4&s=5&s=6&s=7\\
    \hline
    k=1&95.4 &95.7 &95.9 &96.1 &96.1 &96.0 &95.8\\
    \hline
    k=2&96.3 &96.4 &96.6 &96.8 &96.8 &96.7 &96.4\\
    \hline
    k=3&97.2 &97.2 &97.3 &97.5 &97.4 &97.3 &97.3\\
    \hline
    k=4&97.4 &97.6 &97.8 &\bf{97.9} &\bf{97.9} &97.7 &97.5\\
    \hline
    k=5&97.2 &97.3 &97.3 &97.4 &97.5 &97.5 &97.3\\
    \hline
    k=6&96.8 &97.0 &97.0 &97.1 &97.0 &96.9 &96.9\\
    \hline
    \end{tabular}}
\end{table}
%%%%%%%%%%%%%%% end table 5 %%%%%%%%%%%%%%%%%%%%%%%%

{\bf Effectiveness of DeIM}: We utilize t-SNE method~\cite{donahue2014decaf} to visualize the latent features, presented in Fig.~\ref{fig_tSNE_all}, which are generated by the bottleneck layers of the baseline model, i.e., ResNeXt, RCoNet$^k_s$ and three variants of RCoNet$^k_s$: (a) RCoNet-D: a model contains only DeIM; (b) RCoNet-M: a model contains only MUL; (c) RCoNet-DM: a model contains DeIM and MUL but not MHMF. Comparing the latent feature distribution by the baseline model shown in Fig.~\ref{tSNE1:a}, and that by RCoNet-D presented in Fig.~\ref{tSNE1:b}, we can tell that the introduction of DeIM leads to better class separation in the latent space. 

{\bf Effectiveness of MHMF}: We can observe in Fig.~\ref{tSNE1:a} - Fig.~\ref{tSNE1:d} that the latent features of the COVID-19 samples, generated by the models without MHMF, always distribute around the category boundary, and are not quite separable from those of some pneumonia samples. Meanwhile, the latent feature distributions presented in Fig.~\ref{tSNE1:e} - Fig.~\ref{tSNE1:h} derived by the models with MHMF show significant separability between different categories, which implies that MHMF can extract discriminative features.  
We also include numerical results of RCoNet$^k_s$, trained and tested on COVIDx dataset, with regards to different values of $k$, i.e., the number of levels of the moment features to be mixed, and $s$, i.e., the number of experts, in Table~\ref{table_ablation} in terms of accuracy. We can observe that, for a given value of $s$, the accuracy increases first with the value of $k$ but decreases after $k$ is larger than $4$. 
It demonstrates that including more levels of moment feature could improve the model performance. However, the overly high-order moments may lead to performance degradation, which may be because these features are not useful for COVID detection. 

{\bf Effectiveness of MUL}: From Table~\ref{table_ablation}, we observe that, for a given value of $k$, accuracy increases first with the value of $s$ but saturates around $s=5$.
This implies that having more experts in MUL can increase the prediction accuracy but it is not necessary to have too many.

\begin{figure}[t]
    \centering 
    \includegraphics[width=7.9cm]{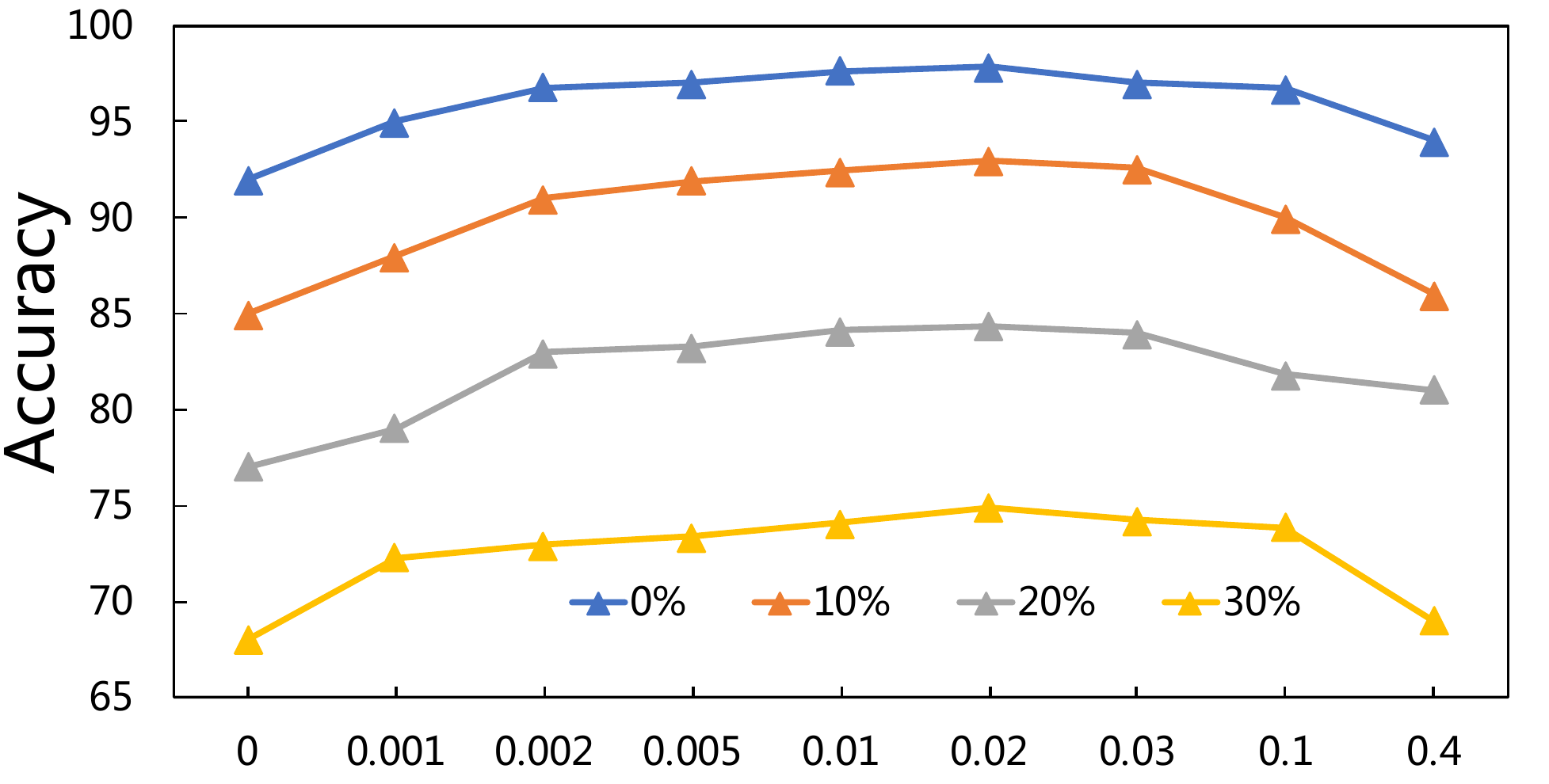}
    \caption{The prediction accuracy by RCoNet$^4_4$ with regards to different values of $\alpha$.}
    \label{hyper}
\end{figure}

\begin{figure}[tb]
    \centering 
    \includegraphics[width=8cm]{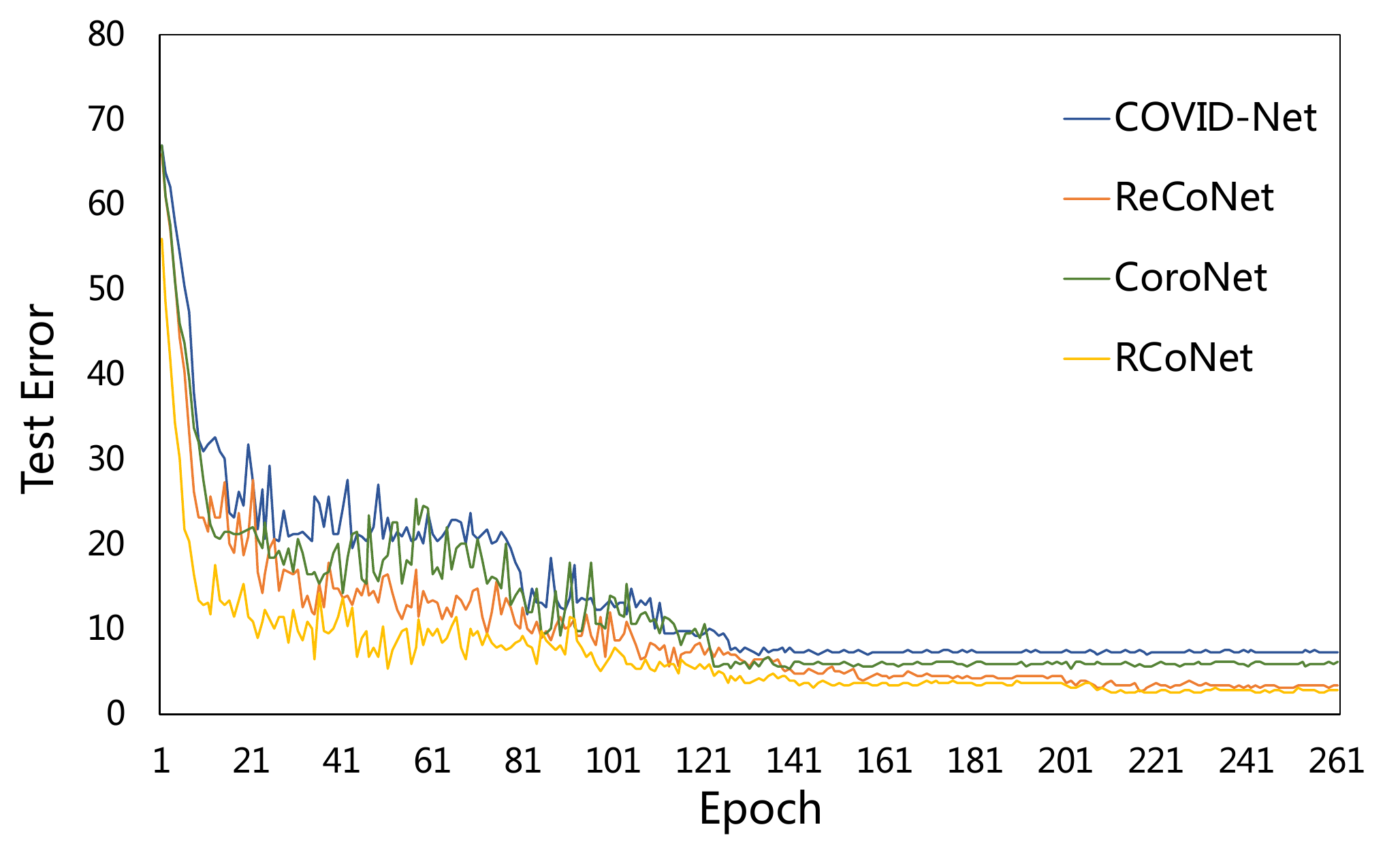}
    \caption{Comparison on the learning trajectories of different models.}
    \label{convergence}
\end{figure}

{\bf Parameter Sensitivity and Convergence}: 
We evaluate how sensitive the model performance in terms of accuracy to the value of $\alpha$. We show the average accuracy of five independent experiments by RCoNet$^4_4$ trained on the dataset with different ratios of noisy samples in Fig.~\ref{hyper}. As we can see, the larger $\alpha$, which means the prediction loss, i.e., $\mathcal{L}_M$, contributes less to the total loss, not necessarily leads to degradation in the accuracy. This means maximizing the mutual information between the input and the latent features could keep useful information within the latent features, thus improving the prediction accuracy. We have also shown the learning curves of different models in Fig.~\ref{convergence}, which shows that RCoNet$^4_4$ converges slightly faster than the others, including COVID-Net, ReCoNet and CoroNet.
%%%%%%%%%%%%%%% begin figure 8 %%%%%%%%%%%%%%%%%%%%%%%%
%%%%%%%%%%%%%%% end figure 8 %%%%%%%%%%%%%%%%%%%%%%%%
%%%%%%%%%%%%%%% begin figure 9 %%%%%%%%%%%%%%%%%%%%%%%%
%%%%%%%%%%%%%%% end figure 9 %%%%%%%%%%%%%%%%%%%%%%%%